\documentclass[11pt,a4paper]{article}
\pdfoutput=1
\usepackage{jcappub}
\usepackage{lineno}
\usepackage{amsmath, amssymb}
\usepackage{bbold}
\usepackage{marginnote}
\usepackage{placeins}
\usepackage[normalem]{ulem}
\usepackage{epsfig}
\usepackage{bm}
\usepackage{tabu}
\usepackage{boldline}
\usepackage{xspace}
\usepackage{slashed}
\usepackage{multirow}
\usepackage{diagbox}
\usepackage[title]{appendix}
\usepackage{subcaption}
\usepackage[table]{xcolor}
\usepackage{enumitem}
\usepackage[utf8]{inputenc}
\usepackage{colortbl}
\usepackage{multirow}
\usepackage[normalem]{ulem}
\usepackage{setspace}
\usepackage{float}
\usepackage{bbm}
\usepackage{grffile}
\usepackage{soul}

\usepackage{cellspace}
\definecolor{orange}{rgb}{1,0.5,0}
\definecolor{blue-violet}{rgb}{0.54, 0.17, 0.89}
\definecolor{rose}{rgb}{0.72, 0.43, 0.47}
\definecolor{UOgreen}{cmyk}{.96,.26,1.00,.15}
\hypersetup{colorlinks,bookmarksopen,bookmarksnumbered,
linkcolor=UOgreen,urlcolor=black,citecolor=UOgreen}

\newcommand{\vmin}{v_{\rm min}}
\newcommand{\vesc}{v_{\rm esc}}
\newcommand{\rhoDM}{\rho_{\rm \chi}}
\newcommand{\mDM}{m_{\rm \chi}}

\newcommand{\NE}{e^-}
\newcommand{\sige}{\overline\sigma_e}
\newcommand{\FDM}{F_{\rm DM}}
\newcommand{\beq}{\begin{equation}}
\newcommand{\eeq}{\end{equation}}
\newcommand{\be}{\begin{equation}}
\newcommand{\ee}{\end{equation}}

\def\GeV{\,{\rm GeV}}

\def\MeV{\,{\rm MeV}}

\def\eV{\,{\rm eV}}
\def\ltap{\ \raise.3ex\hbox{$<$\kern-.75em\lower1ex\hbox{$\sim$}}\ }
\def\gtap{\ \raise.3ex\hbox{$>$\kern-.75em\lower1ex\hbox{$\sim$}}\ }
\def\lsim{\ \raise.3ex\hbox{$<$\kern-.75em\lower1ex\hbox{$\sim$}}\ }
\def\gsim{\ \raise.3ex\hbox{$>$\kern-.75em\lower1ex\hbox{$\sim$}}\ }
\newcommand{\eg}{{\it e.g.}}
\newcommand{\ie}{{\it i.e.}}

\DeclareMathOperator\erf{erf}

\graphicspath{{figs/}}

\begin{document}
\hspace*{110mm}{\small \tt FERMILAB-PUB-23-612-T}
\title{
 Extracting Halo Independent Information from Dark Matter Electron Scattering Data
}
\author[a]{Elias Bernreuther,}
\emailAdd{ebernreu@fnal.gov}
\author[a]{Patrick J. Fox,}
\emailAdd{pjfox@fnal.gov}
\author[b,c]{Benjamin Lillard,}
\emailAdd{blillard@uoregon.edu}
\author[c]{Anna-Maria Taki,}
\emailAdd{annamariataki@gmail.com}
\author[b,c]{and Tien-Tien Yu}
\emailAdd{tientien@uoregon.edu}
\affiliation[a]{Particle Theory Department, Theory Division, Fermilab, Batavia, IL 60510, USA}
\affiliation[b]{Department of Physics, University of Oregon, Eugene, OR 97403, USA}
\affiliation[c]{Institute for Fundamental Science, University of Oregon, Eugene, OR 97403, USA}

\date{\today}


\abstract{Direct detection experiments and the interpretation of their results are sensitive to the velocity structure of the dark matter in our galactic halo. In this work, we extend the formalism that deals with such astrophysics-driven uncertainties, originally introduced in the context of dark-matter-nuclear scattering, to include dark-matter-electron scattering interactions. Using mock data, we demonstrate that the ability to determine the correct dark matter mass and velocity distribution is depleted for recoil spectra which only populate a few low-lying bins, such as models involving a light mediator. We also demonstrate how this formalism allows one to test the compatibility of existing experimental data sets (\eg\ SENSEI and EDELWEISS), as well as make predictions for possible future experiments (\eg\ GaAs-based detectors).  
}

\maketitle
\section{Introduction}

Despite considerable progress on several fronts, and a concerted effort experimentally and theoretically, the properties of particle dark matter (DM) are still unknown. One of the key tools in the hunt for DM is direct detection experiments. These experiments search for the signs of DM in our galactic halo scattering off Standard Model (SM) particles in the (typically underground) lab. Motivated by the WIMP paradigm, these experiments originally focused on DM scattering off nuclei and were best suited for DM masses above $\sim 1~\GeV$. More recently there has been a surge of interest and progress in DM-electron scattering experiments, to search for sub-GeV dark matter~\cite{Essig:2022dfa}.

The expected scattering rate at direct detection experiments depends upon details of the target material, the nature of the DM-SM couplings, and the local DM distribution. The first of these, while challenging to compute, can be determined from experiment and theory. The second is presently unknown but at the low energies involved in the scattering there are only a finite number of possibilities to explore. The last unknown consists of the DM's distribution in both position and velocity space. Typical distributions can be determined from numerical simulations~\cite{diemand2008clumps, klypin2011dark, guedes2011forming, hopkins2018fire} and in our own galaxy the average DM density at the Sun's galactic radius is constrained by observations of stellar kinematics~\cite{Bovy:2012tw}. However, neither approach has the resolution to determine the DM distribution in our local neighborhood. When interpreting results from direct detection experiments, it is typically assumed that the speed of the DM impinging on the detector follows a Maxwell-Boltzmann distribution.  

For DM scattering off nuclei, techniques have been developed to allow interpretation of direct detection results without needing to make this assumption~\cite{Fox:2010bu,Fox:2010bz,Gondolo:2012rs,DelNobile:2013cta,Feldstein:2014gza,Fox:2014kua,Feldstein:2014ufa,Anderson:2015xaa} and thus circumventing the large uncertainty on the DM distribution. This is done by noticing that for nuclear scattering the rate factorizes into a product of a target-dependent term and a target-independent, but halo-dependent, term. This then allows the extraction of (integrals of) the halo velocity distribution from experimental data and the direct comparison of two direct detection rates measured, or bounded, in different materials.  

As with nuclear recoil experiments, astrophysical uncertainties have a significant effect on the interpretation of DM-electron scattering data~\cite{Radick:2020qip,Maity:2020wic}. For DM-electron scattering the target electron's momentum is unknown and must be integrated over. For a given observed recoil energy the range of possible electron momenta depends upon the incoming DM velocity. This means that the scattering rate no longer cleanly factorizes. Instead, the rate involves a convolution of the electron wavefunction with the halo velocity distribution. However, it is still possible to extend halo-independent techniques to the case of electron scattering~\cite{Chen:2022xzi,Chen:2021qao}.

Heuristically, the dependence of the scattering rate for nuclear recoils on astrophysics, \ie\ the DM velocity distribution in the galactic frame $f_\chi$, is 
\be
\frac{d R_n}{dE} \propto \int_{v_{\mathrm{min}}^{\mathrm{nuc}}(E)}^{v_{\mathrm{max}}}d^3v \frac{f_{\chi}(\vec{v}+\vec{v}_E)}{v} ~,
\ee
with $v_{\mathrm{max}}$ the maximal DM speed in the Earth's frame and we have suppressed the terms independent of astrophysics.  There is a one-to-one correspondence between the nuclear recoil energy $E_R$ and the minimum necessary incoming DM speed
$
v_{\mathrm{min}}^{\mathrm{nuc}} = \sqrt{\frac{m_N E_R}{2\mu_{\chi N}^2}}
$,
with $m_N$ denoting the mass of the nucleus and $\mu_{\chi N}$ the reduced mass of the DM-nucleus system.
Thus, the results of different experiments can be compared by analyzing their results in $v_{\mathrm{min}}$-space.

This situation is to be contrasted to electron recoils where the velocity distribution enters in a convolution,
\be
\frac{d R_e}{dE} \propto \int dq \int_{v_{\mathrm{min}}(E,q)}^{v_{\mathrm{max}}}d^3v \frac{f_{\chi}(\vec{v}+\vec{v}_E)}{v} ~,
\ee
with $v_{\mathrm{min}}$ given below, in (\ref{eq:vmin}).

In section~\ref{sec:formalism} we describe the formalism and our statistical approach to finding the best-fit DM mass and velocity distributions.  In this work, we focus on semiconductor targets, but our approach can be readily applied to other detector materials. In section~\ref{sec:mockdata} we apply the method to mock data generated for several different DM models. Our ability to uncover the DM parameters used to generate the mock data improves with the number of populated electron-hole bins. In section~\ref{sec:senseivsedelweiss} we use our approach to compare the consistency of results from SENSEI and EDELWEISS, concluding that their excesses cannot be simultaneously explained as coming entirely from dark matter events.  In addition, we predict the rate in a future GaAs detector under the assumption that the SENSEI observations are due to dark matter. We conclude in section~\ref{sec:conclusions}.

\section{Formalism}
\label{sec:formalism}
In this section we present the method for calculating the scattering rate of DM off of electrons for an arbitrary velocity distribution. A general velocity distribution can be thought of as a sum of DM streams.  Following the approach previously applied to nuclear recoils~\cite{Fox:2010bu, Fox:2010bz, Gondolo:2012rs,DelNobile:2013cta,Feldstein:2014gza,Fox:2014kua,Feldstein:2014ufa,Anderson:2015xaa}, we show how an observed signal can be used to extract the parameters associated with these streams.  We discuss the statistical techniques we use to do this, along with their associated uncertainties.

\subsection{Dark-Matter-Electron Scattering Rate}

To begin, we review the calculations behind the DM-electron scattering rate. 
The general form for the differential scattering rate is given by~\cite{Essig:2015cda}
\beq
\frac{dR}{d \ln E_{e}} = \frac{\rhoDM}{m_{\chi}}\frac{\sige}{8\mu^{2}_{\chi e}}\int  dq\, q\,|\FDM(q)|^{2}  \vert  f_{\rm res}(E_e, q)\vert^{2} \eta \left(\vmin\right)
\label{eq:rate}
\eeq
where $E_e$ and $q$ are the final electron's energy and the momentum transfer, respectively. $m_\chi$ is the DM mass, while $\rhoDM=0.4$ GeV/cm${^3}$ is the local DM density. $\mu_{\chi e}$ is the reduced mass of the DM-e system. $\sige$ denotes the DM-free electron scattering cross section at fixed momentum transfer $q_0=\alpha m_e$, where $m_e$ is the electron mass. $\FDM(q)=(\alpha m_e/q)^n$ encodes the momentum dependence of the interaction, where $n=0\,(2)$ corresponds to a heavy (light) mediator. We assume that the interaction cross section does not directly depend on the DM velocity. $f_{\rm res}(E_e, q)$ is the material-dependent, dimensionless response function for an electron excitation with momentum $q$ and energy $E_e$. 
For a crystal target, we have,
\beq
\label{eq:fcrystal}
|f_{\rm res}(E_e, q)|^2 \equiv \frac{8\alpha m_e^2 E_e}{q^3}\times |f_{\rm crystal}(E_e,q)|^2\, ,
\eeq
where $\alpha\simeq1/137$ is the fine-structure constant and $f_{\rm crystal}(E_e,q)$ is the dimensionless crystal form factor\footnote{For an atomic target, 
$
|f_{\rm res}(E_e,q)|^2 \equiv |f_{nl}^{\rm ion}(E_e,q)|^2\, ,
$
where $(n,l)$ are the quantum numbers for the initial-state bound electron~\cite{Essig:2012yx,Essig:2017kqs}.} as defined in~\cite{Essig:2015cda}. 
Lastly, the DM velocity distribution in the galactic frame, $f_{\chi}(\vec{v})$, enters through $\eta(\vmin)$, 
\beq
\label{eq:etadefinition}
\eta(\vmin)=\int_{\vmin} d^3v \frac{f_{\chi}(\vec{v}+\vec{v}_E)}{v}\, ,
\eeq
where $\vec{v}_E$ is the Earth’s velocity, also in the galactic frame, and we have ignored the small amount of annual time dependence induced by the Earth's orbit around the Sun.
For DM-electron scattering the minimum speed for scattering is 
\beq
\vmin(q) = \frac{q}{2\mDM}+\frac{E_e}{q}\, .
\label{eq:vmin}
\eeq
Thus, in contrast to the DM-nuclear case where $\vmin^{\rm nucl} = \sqrt{\frac{m_NE_R}{2\mu_{\chi N}^2}}$, there is not a clear mapping between recoil energy and $\vmin$, as  (\ref{eq:vmin}) also depends on $q$. Instead, a fixed $E_e$ leads to a range of $\vmin$ as a function of $q$. 
Alternatively, for a fixed $v$ and $E_e$, $q$ must lie in the range $q_- \le q \le q_+$ with 
\beq
q_\pm(v)=\mDM v \pm\sqrt{\mDM^2 v^2-2E_e\mDM}~.
\label{eq:qpm}
\eeq
Furthermore, at fixed recoil energy there is a smallest possible speed allowed for scattering, and an associated momentum transfer to the electron,
\be
\label{eq:v_star}
v_*^2 = \frac{2E_e}{m_\chi}\quad \mathrm{and} \quad q_* = m_\chi v_*~.
\ee

\begin{figure}[t]
\begin{center}
\includegraphics[width=0.49\textwidth]{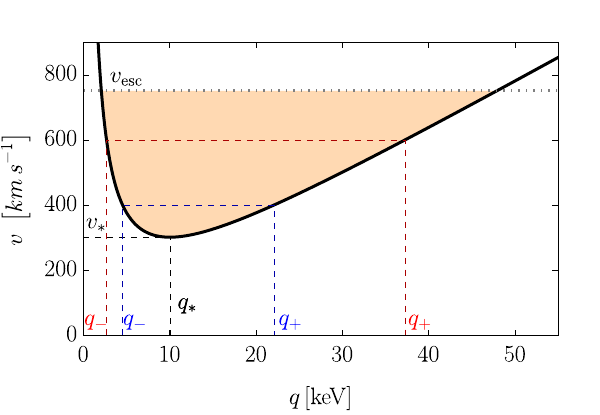}
\includegraphics[width=0.49\textwidth]{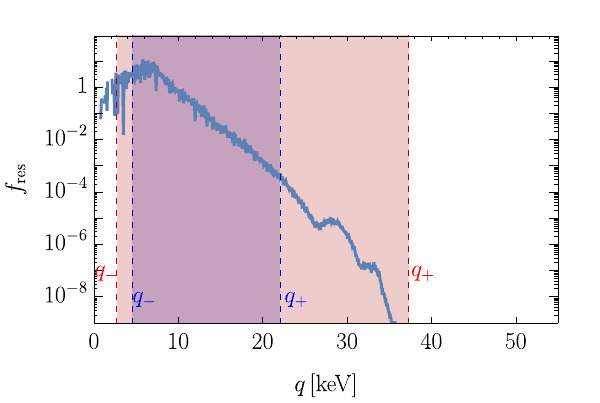}
\end{center}
\caption{{\bf Left:} the region in $q-v$ space that must be integrated over (shaded) for a $10\MeV$ DM candidate scattering off germanium, depositing $5.05\eV$ energy.  The black line is $\vmin$ as a function of momentum exchange $q$. At fixed $q$ the integral over $v$ is from $\vmin$ to $v_{\mathrm{esc}}$, denoted by the dotted line.  At fixed $v$ the integral is over $q_- \le q \le q_+$, denoted by the dashed red/blue lines. The smallest SM speed probed and its corresponding momentum exchange are marked with $v_*, q_*$, respectively. {\bf Right:} The dimensionless response function $f_{\mathrm{res}}$ and the ranges of $q$ probed for the two choices of $v$ in the left panel.}
\label{fig:integrationregion}
\end{figure}

An example of the region in $(q, v)$ space that must be integrated over, at fixed recoil energy, is shown in figure~\ref{fig:integrationregion}.  Note that the response function $f_{\mathrm{res}}$ is a relatively peaked function.  In the limit that  $f_{\mathrm{res}}(q)\rightarrow \delta(q-q_0)$ the relationship between $\vmin$ and the recoil energy becomes one-to-one.  In this case the behavior is much like for nuclear recoils where experiments that probe the same range of $\vmin$-space provide cross-checks of one another.  The ability to directly compare the results of different experiments depends upon the  details of the broader (physical) response functions.  Furthermore, the relationship between recoil energy and $\vmin$ (\ref{eq:vmin}) becomes independent of DM mass when this mass is large and thus it is not possible to separate between different mass hypotheses in this regime.

Typically, one imposes that the velocity distribution of the DM halo, $f_{\chi}(v)$, follows a standard Maxwell-Boltzmann form, which depends on the circular velocity $v_0$ and galactic escape velocity $\vesc$. However, as demonstrated in~\cite{Radick:2020qip}, the DM-electron scattering rate and, accordingly, the cross-section constraints and projections are highly sensitive to the choice of underlying halo model and halo parameters. Therefore, we propose a method in which one can interpret the results of DM-electron scattering experiments in a halo-independent manner.

\subsection{Halo-Independent Method}
We begin by parameterizing the rate as follows,
\beq
\frac{dR}{d \ln E_{e}} \equiv \frac{1}{8\mu^{2}_{\chi e}}\int  dq\, q\, |\FDM(q)|^{2}  \vert  f_{\rm res}(E_e, q)\vert^{2} \tilde\eta \left(\vmin\right)
\label{eq:ionization_tilde}
\eeq
where 
\beq
\tilde\eta(\vmin)=\frac{\rhoDM \sige}{\mDM}\eta(\vmin)\, .
\eeq
Since $\eta(\vmin)$ is given as the integral (\ref{eq:etadefinition}) over the velocity distribution, $\eta(\vmin)$ and, therefore, $\tilde\eta(\vmin)$ are monotonically-decreasing functions of $\vmin$. We adopt a conservative ansatz by approximating $\tilde{\eta}(\vmin)$ as a series of step functions~\cite{Feldstein:2014gza,Fox:2014kua,Feldstein:2014ufa}. 
The intervals are defined by dividing up $\vmin$-space into $N_s$ intervals of the form $[v_i, v_{i+1}]$, with $v_i \le v_{i+1}$.  This corresponds to the scenario in which the velocity distribution of the DM halo is given by a series of streams with speeds $\{v_i\}$ and thus  
\beq
\label{eq:etatilde_ansatz}
\tilde\eta(\vmin)=\sum_i \tilde\eta_i\,\Theta(v_i-\vmin)\, .
\eeq
Note that, by construction, $\tilde\eta$ is monotonically decreasing with $\vmin$. As $\vmin$ decreases, the step across speed $v_i$ is of height $\tilde{\eta}_i$.
We discuss the technical details of our numerical procedure below, see section \ref{subsec:stats}.

For several sub-GeV DM experiments, the observed event rates are given as a function of the number of detected electrons, $n_e$. We use a simple model to determine the correspondence between the electron energy and $n_e$, given by 
\beq
n_e=1+\lfloor{(E_e-E_{\rm gap})/\varepsilon}\rfloor\, ,
\eeq
where $\lfloor x\rfloor$ rounds x down to the nearest integer, $E_{\rm gap}$ is the bandgap energy of the material, and $\varepsilon$ is the mean energy per electron-hole pair. For silicon, $\{E_{\rm gap},\varepsilon\}=\{1.2,3.8\}$ eV, while for germanium $\{E_{\rm gap},\varepsilon\}=\{0.67,3.0\}$ eV.  The observed rate for $n_e$ detected electrons is given by integrating the differential rate (\ref{eq:ionization_tilde}) over the energies associated with that number of electrons,
\beq
R(n_e) = \frac{1}{8\mu^{2}_{\chi e}}\int_{E_{\rm min}}^{E_{\rm max}} \frac{dE_e}{E_e}\int_{q_-}^{q_+}  dq\, q\, |\FDM(q)|^{2}  \vert  f_{\rm res}(E_e, q)\vert^{2} \sum_i \tilde\eta_i\Theta(v_i-\vmin)\, ,
\label{eq:rate_ne}
\eeq
where ${E_{\rm min}} = E_{\rm gap}+(n_e-1)\varepsilon$ and ${E_{\rm max}} = E_{\rm gap}+n_e\varepsilon$. The limits of $q$ are determined by kinematics, and are given by $q_\pm$ in (\ref{eq:qpm}).

\subsection{Statistics}
\label{subsec:stats}
In order to determine the constants $\tilde\eta_i$, we  introduce a likelihood function which compares the observed number of events to the number predicted for a given choice of $\tilde\eta_i$. For our purposes, we will use a binned likelihood function for an experiment with $N$ bins,
\beq
-2\log{\cal L}=2\sum_{i=1}^{N}\left[R_i+B_i-N_i+N_i\log\frac{N_i}{R_i+B_i}\right]\, ,
\label{eq:minLL}
\eeq
where $R_i,\,B_i,\,N_i$ are the expected signal, expected background, and the observed number of events for the $i$-th bin of the experiment, respectively. The expected signal rate $R_i$ in each bin is determined using (\ref{eq:rate_ne}), and any known sources of background can be accounted for with $B_i$.
For the case where $N_i=0$, we set $\left[N_i\log\frac{N_i}{R_i+B_i}\right]=0$. For more than one experiment we label each likelihood with an index $\alpha$, and the results of all experiments are combined by multiplying (adding) their (log-) likelihoods, 
\be
\mathcal{L} = \prod_\alpha \mathcal{L}^{(\alpha)}~.
\label{eq:fulllikelihhod}
\ee
Given this likelihood, we can determine, for a fixed $\mDM$,  the step heights $\tilde\eta_i$ which minimize $-2\log{\cal L}$ ({\it i.e.~}maximize the likelihood). From this, we can define the profile likelihood function,
\begin{equation}
\hat{\cal L}(\mDM)=\max_{\mathbf{\tilde\eta}}{\cal L}(\mDM,\mathbf{\tilde\eta})\, .
\label{eq:profileL}
\end{equation}
The linearity of the rate~\eqref{eq:rate_ne} with respect to all $\tilde{\eta}_i$ ensures that any minimum of the likelihood in $\tilde{\eta}$ space is a global minimum. To solve this minimization problem in practice, we discretize the function $\tilde{\eta}(v_\mathrm{min})$ in 100 velocity steps of equal size $\Delta v_\mathrm{min}$ between $v_\mathrm{min}=0$~km/s and $v_\mathrm{min}$=1000~km/s. In the galactic frame the maximal speed of dark matter bound in our halo is expected to be $\mathcal{O}(500-700)\mathrm{km}/\mathrm{s}$.  In the Earth's frame the maximal speed is about $200$~km/s higher.  We consider 1000~km/s a conservative cut off, but it is simple to extend the range to higher speeds.  Concretely, we take the step-function ansatz \eqref{eq:etatilde_ansatz} and write the step heights as
\begin{equation}
    \tilde{\eta}_i \equiv \exp(\tilde{g}_i) \times 1 \,\text{year}^{-1} \, ,
\end{equation}
with real parameters $\tilde{g}_i$, where $i=1\dots100$. This parameterization ensures that the step heights are non-negative, as dictated by the monotonicity of $\tilde{\eta}(v_\mathrm{min})$.

To find the values $\tilde{g}_i$ that minimize the log-likelihood \eqref{eq:minLL}, we use the \textsc{ADAM} optimizer \cite{kingma2014adam} implemented in \textsc{TensorFlow}~2.10.0 \cite{tensorflow2015-whitepaper} with its default settings, a learning rate of 0.001, and the log-likelihood as the loss function. After removing an overall normalization of $10^{-20}$~year$^{-1}$, we initialize the $\tilde{g}_i$ to a random value between 0 and 1. We then perform $10^5$ minimization steps, which was sufficient for the optimizer to converge in all cases studied in this work.

While we always use the log-likelihood as a loss function for our optimizer, we will also use the $\chi^2$ as a test statistic, which is given by
\begin{equation}
    \chi^2 = \sum_{i=1}^{N} \frac{\left(N_i-R_i-B_i\right)^2}{R_i+B_i} \, ,
\end{equation}
where all quantities are defined like in  \eqref{eq:minLL}. Compared with the likelihood, the $\chi^2$ has the advantage that it provides an absolute measure of the compatibility of the data with a given model hypothesis.

\subsection{Linear Algebra and Flat Directions}
\label{sec:linalg}

The rate (\ref{eq:rate_ne}) depends linearly on $\tilde \eta_i$. So for fixed dark matter mass the expected rate for a given velocity distribution, corresponding to $\tilde \eta$, is given by
\be
R_i(\tilde\eta) = \sum_j \mathcal A_{ij} \tilde{\eta}_j,
\ee
with
\be
\mathcal A_{i j}(m_\chi) = \frac{1}{8\mu^{2}_{\chi e}} 
\int_{E_{\rm min}^{(i)}}^{E_{\rm max}^{(i)}} \frac{dE_e}{E_e}\int_{q_-}^{q_+}  dq \,q\,|\FDM(q)|^{2}  \vert  f_{\rm res}(E_e, q)\vert^{2} \Theta(v_j-\vmin)\, .
\ee
The index $i$ refers to the different observables, and $\mathcal A(m_\chi)$ encodes how much the $j$th velocity bin contributes to the $i$th observable, assuming some fixed value for $m_\chi$.
If $\mathcal A$ is an invertible square matrix, as would be the case if the number of observed energy bins was equal to the number of steps in the velocity distribution, then a best-fit solution for $\tilde{\eta}_j$ given some observed rates $R_i$ could be found simply by inverting $\mathcal A$:
$ \tilde\eta_j = (\mathcal A^{-1})_{ji} R_i$.

When the number, $n$, of continuous parameters $\tilde{\eta}_i$ is larger than the number of observables, $m$, the matrix $\mathcal{A}$ is not invertible in the usual fashion.  Instead one may introduce the concept of a pseudoinverse\footnote{For a brief introduction on the properties of the pseudoinverse, see appendix~\ref{app:linearalgebra}.} $\mathcal{A}^+$.  The pseudoinverse can be found from the singular value decomposition (SVD) of $\mathcal{A}$.
The SVD identifies orthonormal matrices $U$ and $V$ such that
\begin{align}
\mathcal A = U \mathcal D V^T ~.
\end{align}
If $\mathcal{A}$ is of size $(m\times n)$ then $U(V)$ are square matrices of size $m(n)$, respectively, and $\mathcal D = D \oplus \mathbb{0}$ is of size $(m\times n)$, with $D$ a diagonal $(n \times n)$ matrix and $\mathbb{0}$ a zero matrix.  Defining $\mathcal D^+ = D^{-1} \oplus \mathbb{0}$ the pseudoinverse of $\mathcal{A}$ is defined as
\be
\mathcal{A}^+ = V \mathcal D^+ U^T~.
\ee
Using the pseudoinverse a solution can be found:
\be
\tilde{\eta}_* = \mathcal{A}^+ R~.
\ee
It is shown, in appendix~\ref{app:linearalgebra}, that this solution minimizes
 $|R - \mathcal{A} \tilde{\eta}|^2$.\footnote{By rescaling $R$ and $\tilde{\eta}$ the quantity being minimized can be made into $\chi^2$.
In the limit where the number of events is large -- where the Poisson distribution approaches a Gaussian -- this quantity approaches  $-2 \log \mathcal L$.}  For the case of interest here, with more parameters than constraints, this minimum is at zero.  Thus, $\tilde{\eta}$ is the best-fit solution.
  However, there is no guarantee that the solution is physical \ie\ that the resulting $\tilde{\eta}$ is monotonic.   Furthermore, there are flat directions in the parameter space: perturbations to the vector $\tilde{\eta}$ that leave every observable $R_i$ unchanged.
These flat directions are spanned by the vectors that lie in the kernel of $\mathcal{A}$.  Thus, there is a family of solutions related to $\tilde{\eta}_*$ by
\be
\tilde{\eta}_* \rightarrow \tilde{\eta}_* + \left(\mathbb{1}-\mathcal{A}^+\mathcal{A}\right) \vec{\alpha}~,
\label{eq:flats}
\ee
with $\vec{\alpha}$ an arbitrary vector that is projected down to $\text{ker} (\mathcal A)$.
By moving around in these flat directions an equally good, but physical, solution for $\tilde{\eta}$ may be found.  

Bands of $1\sigma$ or $2\sigma$ can be generated by a related two-step procedure. First, starting from a best-fit point $\tilde{\eta}_*$, the vector $\tilde\eta$ is varied in all of the directions perpendicular to $\text{ker}(\mathcal A)$, until the $\chi^2$ of the fit reaches a $1\sigma$ (or $2\sigma$) threshold. These results for $\tilde{\eta}$ generate a partial contour, surrounding the best-fit point. Next, every point on this contour is extended along the flat directions, using the same method as \eqref{eq:flats}. This complete $1\sigma$ or $2\sigma$ contour encloses a volume, with $\tilde{\eta}_*$ and all of the points generated by \eqref{eq:flats} contained in the center.

With a large number of flat directions, determining in which direction to move to go from the unphysical solution $\tilde{\eta}_*$ to a physical one can be numerically challenging.  However, once a physically allowed solution has been determined, for instance using the techniques outlined in the previous subsection, the pseudoinverse and its associated flat directions can be employed to determine the boundaries of the physically allowed region which yields the same rate as the original solution.  
In high-dimensional applications it may be more tractable to explore a subspace of the flat directions. In section~\ref{sec:mockdata} we employ a coarse-graining method, where $k$ neighboring bins in the distribution are varied in parallel, $\delta \eta_{i} = \delta\eta_{i+1} = \ldots = \delta\eta_{i+k}$. This makes it easier to explore a larger range of the physically-allowed region, or at least those perturbations to $\tilde{\eta}_*$ that do not involve significant changes at the smallest scales. Through this technique, we can then define a Region of Equivalent Statistical Test (REST) that encapsulates the degeneracy in the underlying velocity distribution by containing all velocity distributions that yield the same rate.

\section{Application to simulated rates}
\label{sec:mockdata}
As an application of the formalism described in the previous sections, we simulate mock data under various DM model assumptions and apply the formalism to this data set.  This provides an opportunity to measure the utility of this approach, under controlled conditions.  
We calculate mock data for both silicon and germanium targets.  To compute the rate (\ref{eq:rate}) we use crystal form factors (\ref{eq:fcrystal}) calculated by {\tt QEdark}~\cite{Essig:2015cda}, but the procedure can be applied using rates from other codes, such as {\tt DarkELF}~\cite{Knapen:2021bwg}, {\tt EXCEED-DM}~\cite{Trickle:2022fwt}, and {\tt QCDark}~\cite{Dreyer:2023ovn}.  
We simulated two representative halo models: the Standard Halo Model (SHM) and a stream of DM at a fixed velocity. 

We take the SHM to have the form~\cite{Baxter:2021pqo}
\begin{equation}
\label{eq:shm}
f_\chi(\vec v_\chi)
=\frac{1}{K_{SHM}}e^{-v_\chi^2/v_0^2}\, \Theta(v_{\rm esc} - v_\chi) \, ,
\end{equation}
where $\vec v_\chi$ is the DM velocity in the galactic frame.
We take the velocity dispersion to be $v_0=220$ km/s and $v_{\rm esc}=544$ km/s for the escape velocity.  The distribution is normalized such that $\int d^3v_\chi f_\chi(v_\chi)=1$.  The analytic expression for the normalization factor $K_{SHM}$ is derived in appendix~\ref{app:halomodels}.
To determine the rate in a direct detection experiment we must boost to the Earth's frame.  To do so, we use the average of the Earth's speed relative to the DM halo, $v_E=232$ km/s, and ignore its time dependence.  

In the galactic frame a stream with zero velocity dispersion is a delta function in velocity space, $f_\chi(\vec{v}_\chi) = \delta^{(3)}(\vec{v}_\chi-\vec{v}_{\mathrm{str}}^{\,\mathrm{gal}})$.  Here, we model the DM velocity distribution as a Gaussian with small dispersion, $\sigma$.  Furthermore, we define the stream's mean velocity in the Earth's frame $\vec{v}_{\mathrm{str}} = \vec{v}_{\mathrm{str}}^{\,\mathrm{gal}} - \vec{v}_E$.  Thus, in the Earth's frame the velocity distribution for the stream is
\begin{equation}
\label{eq:stream}
    f(\vec{v})=\frac{1}{\sqrt{8\pi^3}\sigma^3} e^{-\frac{(\vec{v} -\vec{v}_{\mathrm{str}})^2}{2\sigma^2}}\, ,
\end{equation}
where we take the speed of the stream to be ${v}_{\mathrm{str}}=300$ km/s and the dispersion $\sigma=20$ km/s. We are interested in the limit where the stream's dispersion is small and its speed is far from the escape speed, thus we ignore the cutoff at the galactic escape speed. See appendix~\ref{app:halomodels} for more discussion about the function $\eta$ for the cases of a SHM and a stream. 

The DM can couple to electrons either through a contact interaction or through the exchange of a light mediator.  For these two cases we parameterize the DM form factor \cite{Essig:2011nj} as $\FDM=1$ and $\FDM=\left(\alpha m_e/q\right)^2$, respectively.

Using these velocity distributions, the DM form factors, and the crystal form factors (\ref{eq:fcrystal}) calculated by {\tt QEdark}, we simulate four different DM models, whose parameter choices are listed in table~\ref{tab:mockdata}. In this analysis of mock data, for simplicity, we do not include backgrounds and hence always set $B_i=0$. 
\begin{table}[h]
    \centering
    \begin{tabular}{c||c|c|c}
        model name & $m_\chi$ [MeV] & $\FDM$ & halo model \\\hline
        A  & 50 & 1 & SHM\\ 
        B  & 10 & 1 & SHM\\ 
        C  & 50 & $\left(\alpha m_e/q\right)^2$ & SHM\\ 
        D  & 50 & 1 & stream 
    \end{tabular}
    \caption{The DM mass, DM form factor, and DM halo model assumed for each of the generated mock data sets.}
    \label{tab:mockdata}
\end{table}
\subsection{Best-fit dark matter particle and astrophysical properties}

\begin{figure}[tbhp]
\begin{center}
\includegraphics[width=0.495\textwidth]{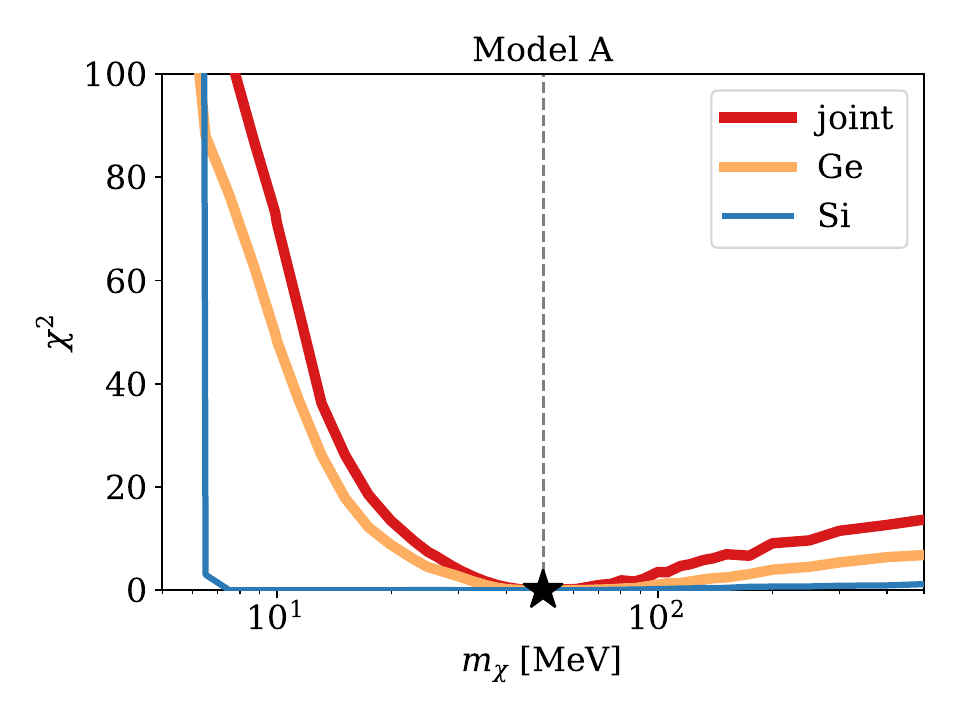}
\includegraphics[width=0.495\textwidth]{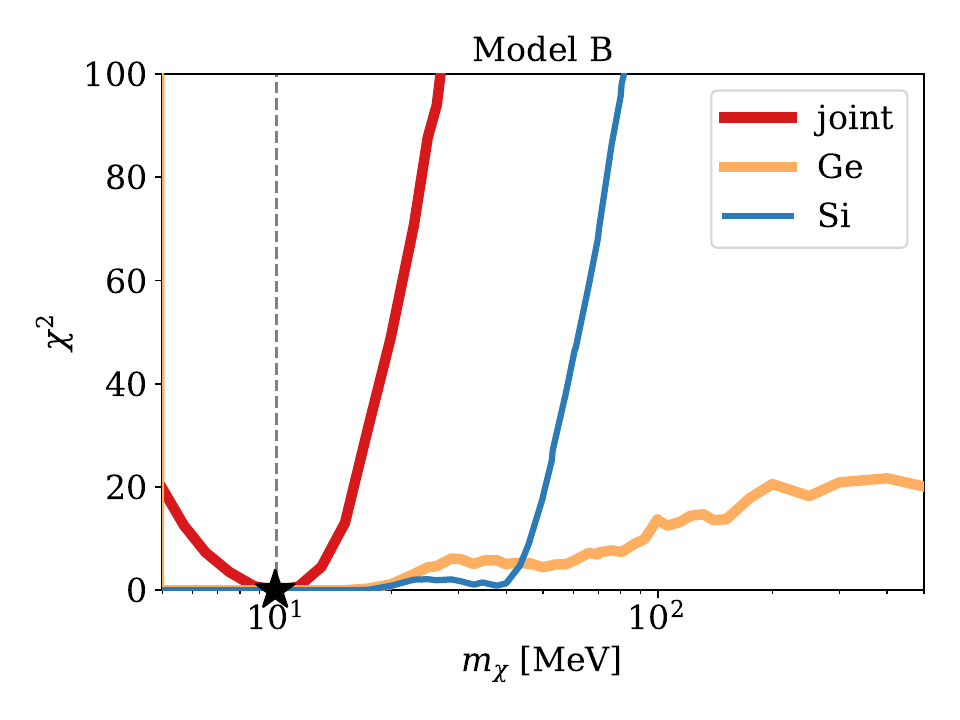}
\includegraphics[width=0.495\textwidth]{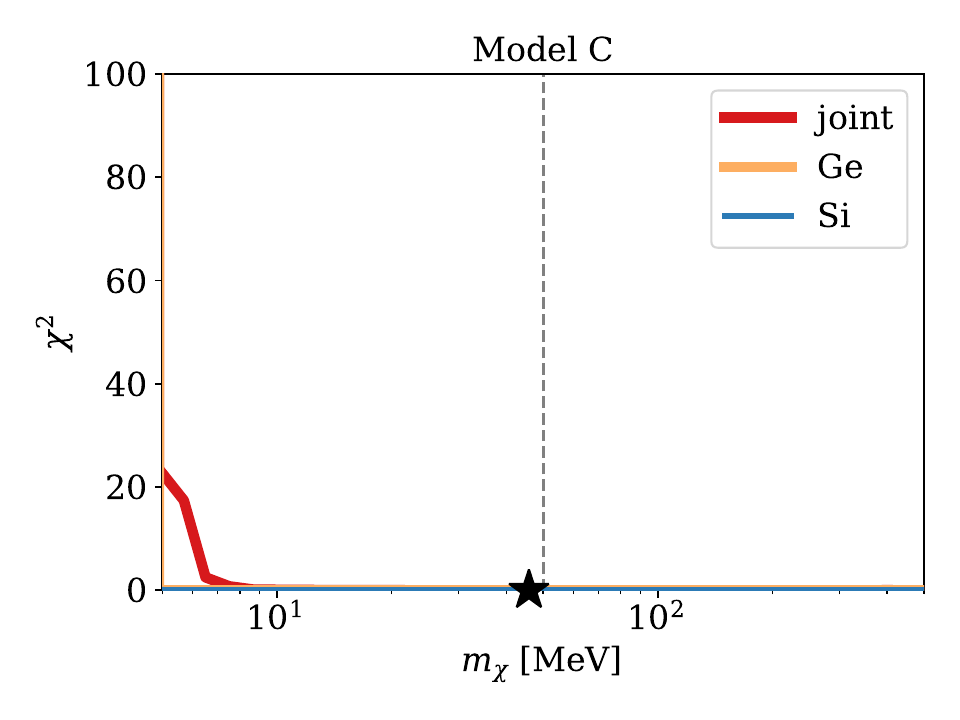}
\includegraphics[width=0.495\textwidth]{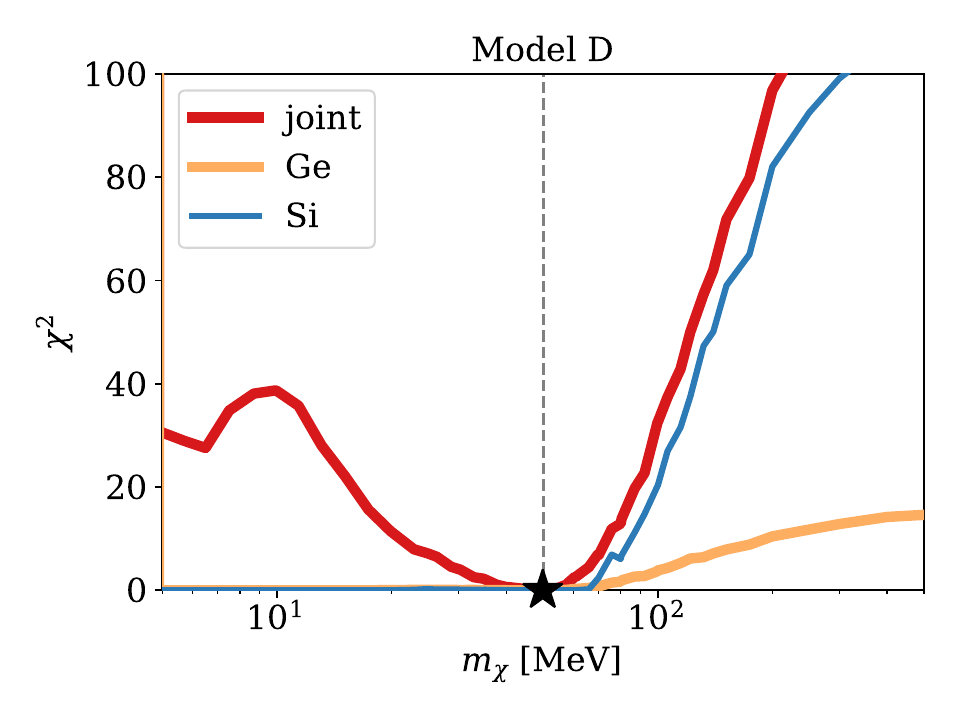}
\caption{Best-fit $\chi^2$ as a function of $m_\chi$ from a fit (profiled over the DM velocity distribution) of simulated Si and Ge rates at 1~kg$\cdot$yr exposure for different DM and halo models. Shown are separate fits of the Si rate (blue) and the Ge rate (orange), and a joint fit of both (red). Each fit is carried out with the correct $F_\mathrm{DM}$ for the respective DM model.  The vertical dashed line shows the true DM mass used in creating the mock data, while $\mathbf{\star}$ 
denotes the DM mass that gives the best joint fit to the two mock data sets. For a version of these plots with a logarithmic $y$-axis, see the dashed curves in Figure~\ref{fig:chisquared_asimov_wrongfdm}.
}
\label{fig:chisquared_asimov}
\end{center}
\end{figure}

\begin{figure}[tbhp]
\begin{center}
\includegraphics[width=0.495\textwidth]{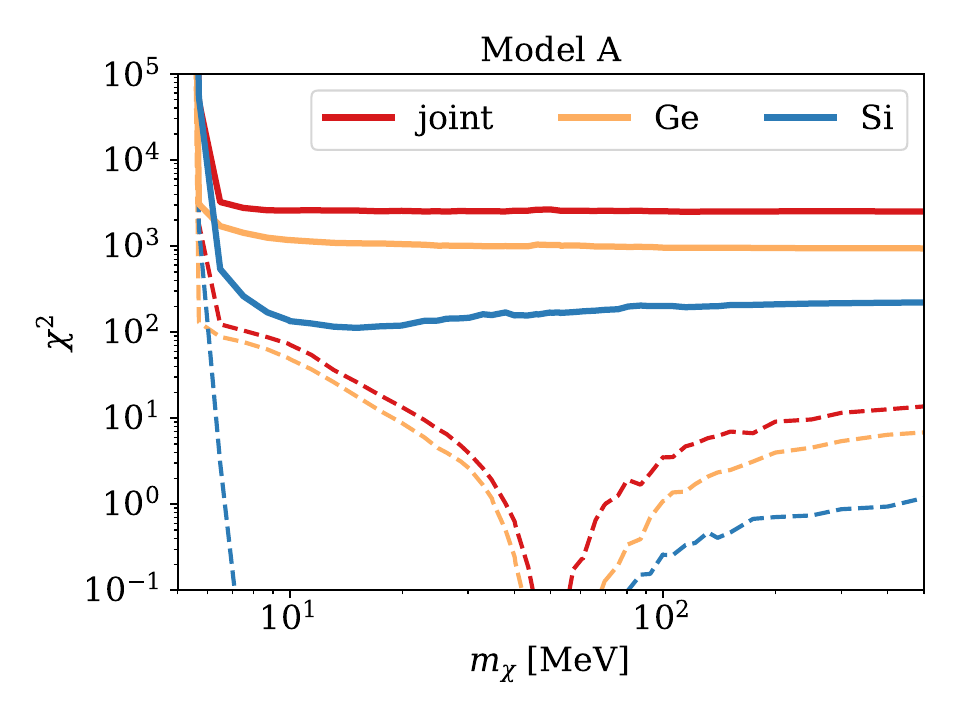}
\includegraphics[width=0.495\textwidth]{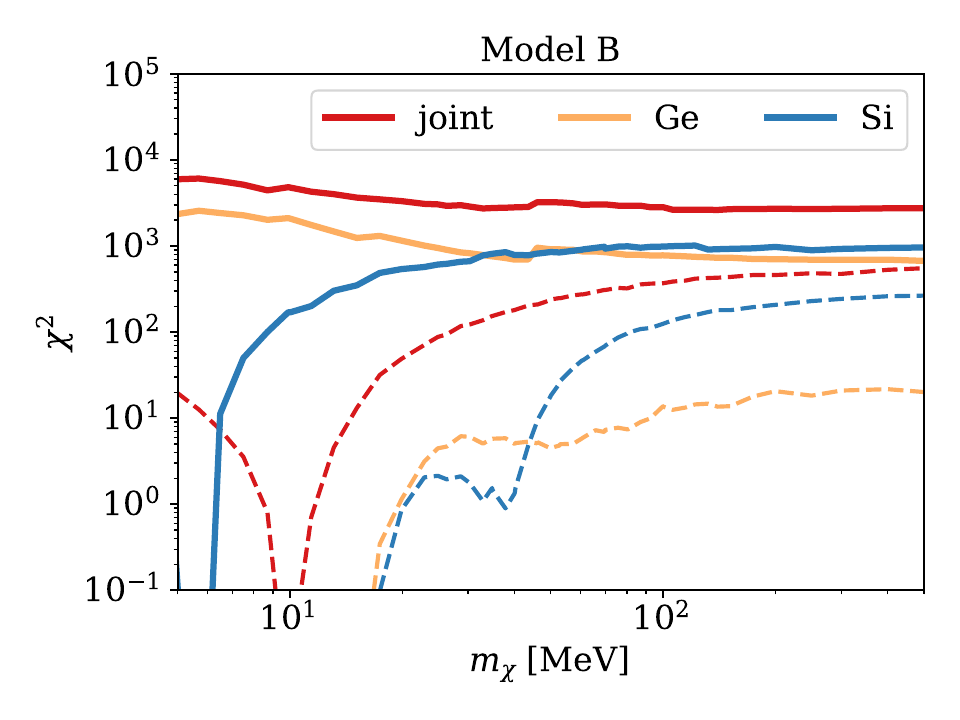}
\includegraphics[width=0.495\textwidth]{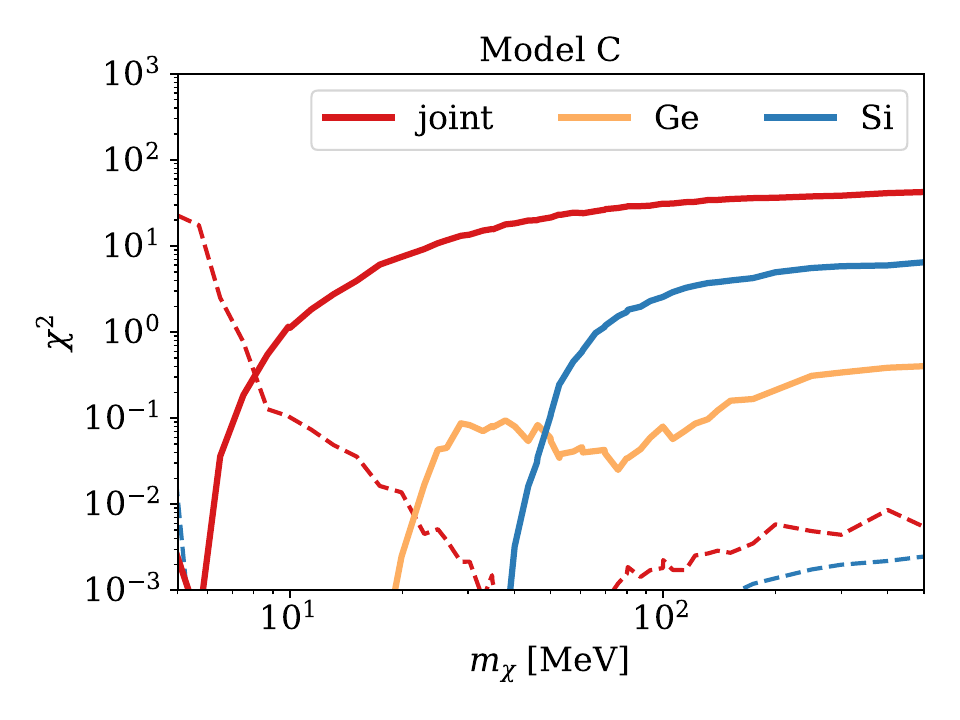}
\includegraphics[width=0.495\textwidth]{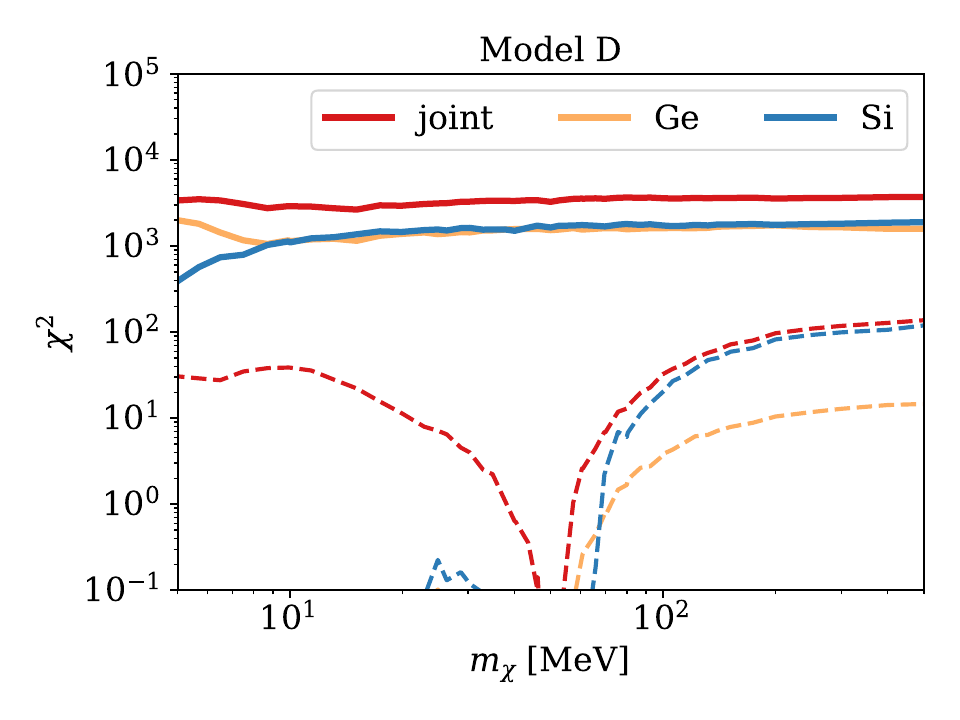}
\caption{Same as figure~\ref{fig:chisquared_asimov}, but with the incorrect DM form factors, i.e.\ $F_\mathrm{DM}\sim 1/q^2$ for models A, B and D, and $F_\mathrm{DM}=1$ for model C.  The results from figure~\ref{fig:chisquared_asimov}, which use the correct form factors, are shown with dashed lines (note the change to a logarithmic scale).
}
\label{fig:chisquared_asimov_wrongfdm}
\end{center}
\end{figure}

Armed with the mock data, we can now employ our formalism to find the best-fit DM velocity distribution for the two elements, silicon and germanium, independently as well as the joint fit of the two rates.  In figures~\ref{fig:chisquared_asimov} and~\ref{fig:chisquared_asimov_wrongfdm}, we show the value of $\chi^2$ profiled over the DM velocity distribution as a function of the hypothesized DM mass for the mock data from models A--D. In each fit of figure~\ref{fig:chisquared_asimov}, the DM form factor $F_\mathrm{DM}$ is set to its true value for the respective model, while in figure~\ref{fig:chisquared_asimov_wrongfdm}, the DM form factor is set to the wrong value.

Using the method described in section~\ref{sec:formalism}, we carry out separate fits of simulated silicon and germanium rates as well as a joint fit of the events from both elements. For the case of the joint fit with the correct $\FDM$, we see that the minimum of the $\chi^2$ statistic, {\it i.e.} the best fit, lies exactly at the true DM mass for models A, B and D, and nearly at the correct mass for model C. In each case, the $\chi^2$ value at the minimum is negligible. That is, for the correct mass we can fit the simulated data perfectly, as expected. At incorrect masses, on the other hand, a perfect joint fit of both silicon and germanium data cannot be achieved with any physical velocity distribution for models A, B and D. This demonstrates that our formalism is successful at identifying the correct DM mass for models with a heavy mediator, {\it i.e.} $F_\mathrm{DM}=1$.

Only for model C, which has a light mediator, can we find some velocity distribution that fits the data perfectly for any DM mass $m_\chi \gtrsim 10$~MeV. Here, the light mediator, corresponding to $F_\mathrm{DM} \sim 1/q^2$, biases the electron recoil spectrum towards small energies largely irrespective of $m_\chi$, erasing information about the DM mass from the resulting event rates. The value of $\chi^2$ only increases at $m_\chi \lesssim 10$~MeV, where fitting the event rates of model C would require velocities above $1000$~km/s, which we impose as a hard cutoff in our fits.

For a heavy DM mediator, our method correctly infers the underlying DM mass via a joint fit of silicon and germanium data. However, when fitting silicon and germanium rates individually, we find in almost every case (the Ge fit of model A being the only exception) a broad range of masses for which some velocity distribution yields negligible $\chi^2$. This demonstrates that it is essential to combine event rates from multiple electron-scattering experiments with different materials to infer DM properties in a halo-independent manner.

Results for incorrect choices of the form factor (\textit{i.e.} $F_\mathrm{DM} \sim 1/q^2$ for models A, B, D, and $F_\mathrm{DM}=1$ for model C) are displayed in figure~\ref{fig:chisquared_asimov_wrongfdm}. As in the previous figure, we again show $\chi^2$ for the best-fit velocity distribution for each DM mass. For comparison, we also show the $\chi^2$ curves with the correct form factor from figure~\ref{fig:chisquared_asimov} as dashed lines.

When fitting silicon and germanium data simultaneously with the incorrect form factor choices, we observe best-fit values of $\chi^2$ between $10^3$ and $10^4$ over the entire mass range for models A, B and D. Only for the light-mediator model C there is a narrow range of masses below approximately 10~MeV where we obtain a good fit despite the wrong DM form factor. This indicates that a light DM particle with a heavy mediator can mimic the event rates of a heavier DM particle with a light mediator. This is not unexpected, since  both scenarios lead to a concentration of events at low energies.

The results for models A, B, and D, on the other hand, show that the combined rates in silicon and germanium from a heavy-mediator model cannot be explained by a light-mediator model with any DM mass or velocity distribution. This implies that our halo-independent analysis method can correctly identify that the coupling is through a heavy mediator even without any input about the DM mass or velocity.

\begin{figure}[tbhp]
	\begin{center}
		\includegraphics[width=0.495\textwidth]{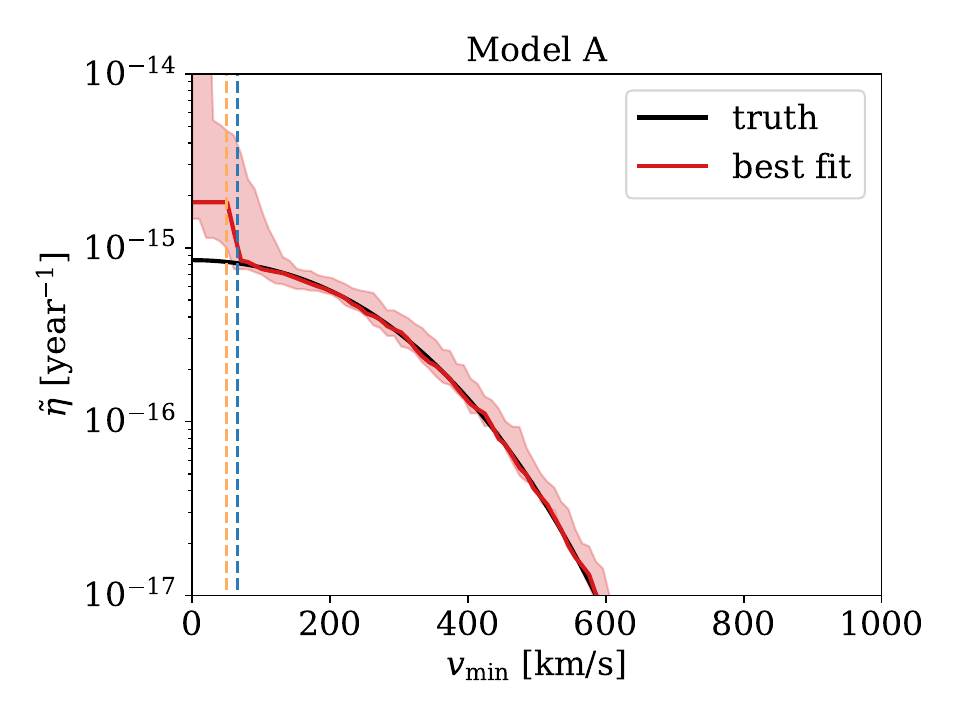}
		\includegraphics[width=0.495\textwidth]{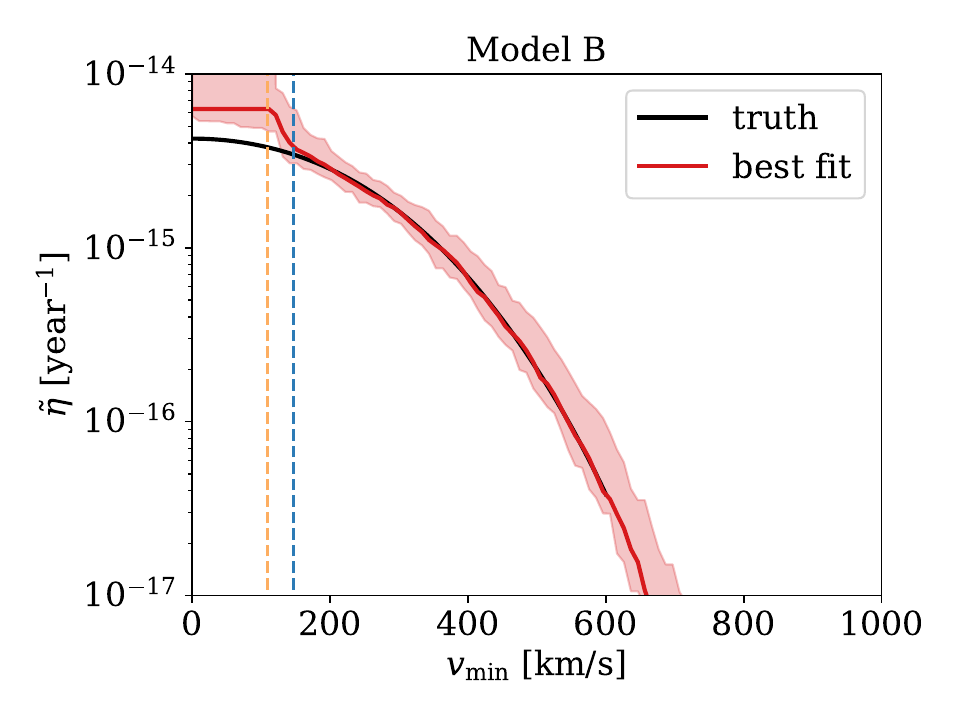}
		\includegraphics[width=0.495\textwidth]{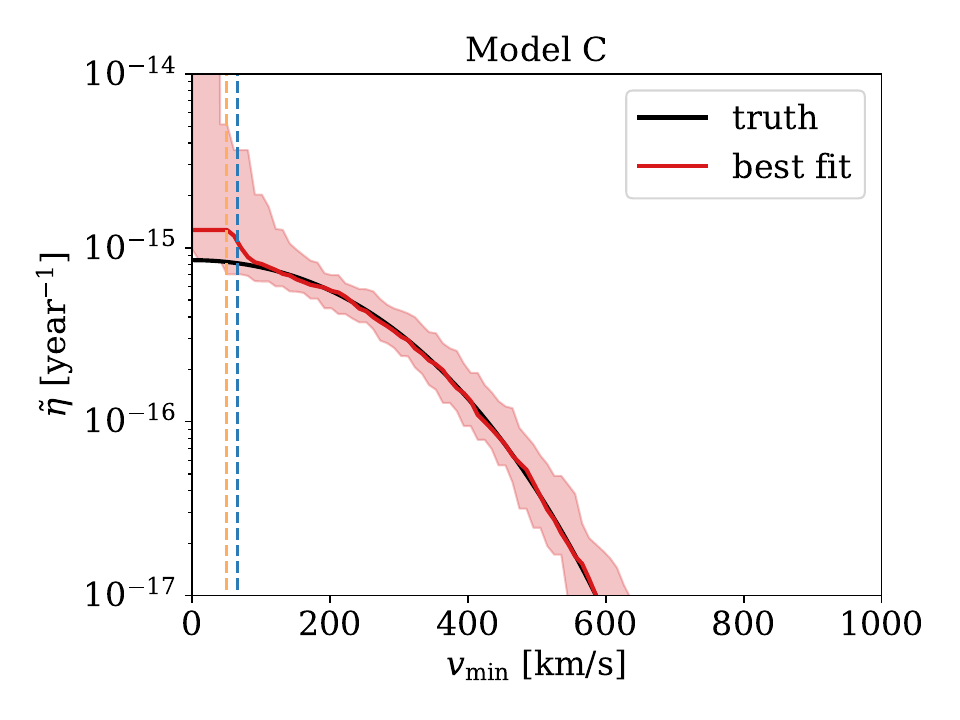}
		\includegraphics[width=0.495\textwidth]{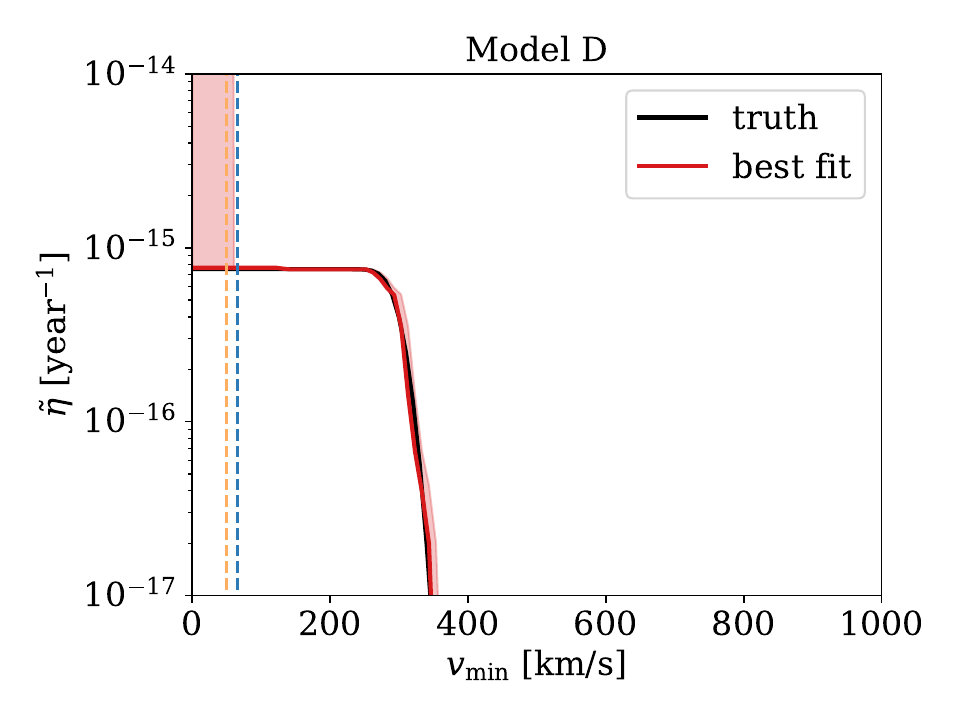}
		\caption{Best-fit velocity distributions, shown in the form of $\tilde{\eta}(\vmin)$, for the correct DM masses and DM form factors for the models from figure~\ref{fig:chisquared_asimov} inferred from a joint fit of simulated Si and Ge rates. The threshold velocities (as given in  \eqref{eq:v_star}) for Si and Ge are shown as blue and orange dashed lines, respectively. The bands are defined by the REST procedure described in section~\ref{sec:linalg}.}
		\label{fig:bestfit_velocity_asimov}
	\end{center}
\end{figure}

Having studied the ability of our formalism to identify the particle properties of dark matter, we next turn to the DM velocity distribution. The best-fit results for $\tilde{\eta}$ for the correct masses of models A--D are shown in figure~\ref{fig:bestfit_velocity_asimov} alongside the $\tilde{\eta}$ of the true velocity distributions for each model. These are the Standard Halo Model defined in  \eqref{eq:shm} for models A--C and a stream localized at $v_\mathrm{str}=300$~km/s for model D, see  \eqref{eq:stream}. Around each best-fit velocity distribution we also show the corresponding REST, which accounts for the flat directions in velocity space, as discussed in section~\ref{sec:linalg}.

For all four models considered here we find that the best-fit velocity distribution follows the true velocity distribution nearly perfectly. Note that the part of the velocity spectrum below the threshold $v_*(E_\mathrm{gap})$ (as defined in  \eqref{eq:v_star}) does not affect the observed event numbers and therefore remains at its arbitrary initial value throughout the minimization process. Hence the best-fit $\tilde{\eta}$ curves reach a plateau at the lowest velocities shown in figure~\ref{fig:bestfit_velocity_asimov}. Accounting for this, the $\tilde{\eta}$ of the true velocity distribution above threshold is contained inside the REST for all models.

\subsection{Limits on the dark matter mass}
\begin{figure}[thbp]
	\begin{center}
		\includegraphics[width=0.495\textwidth]{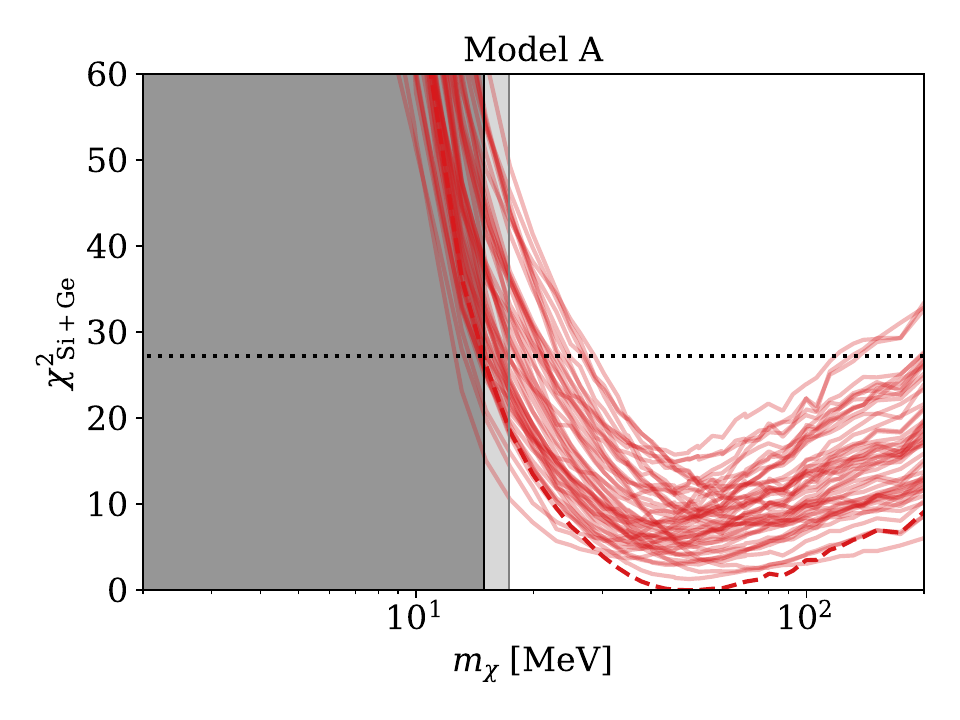}
  		\includegraphics[width=0.495\textwidth]{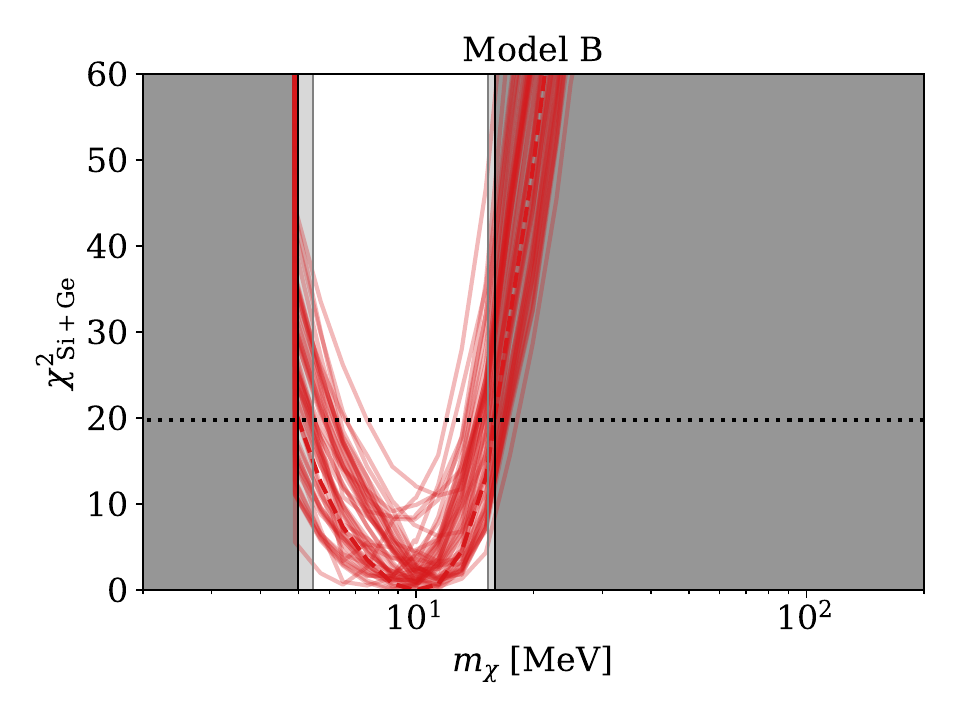}
    	\includegraphics[width=0.495\textwidth]{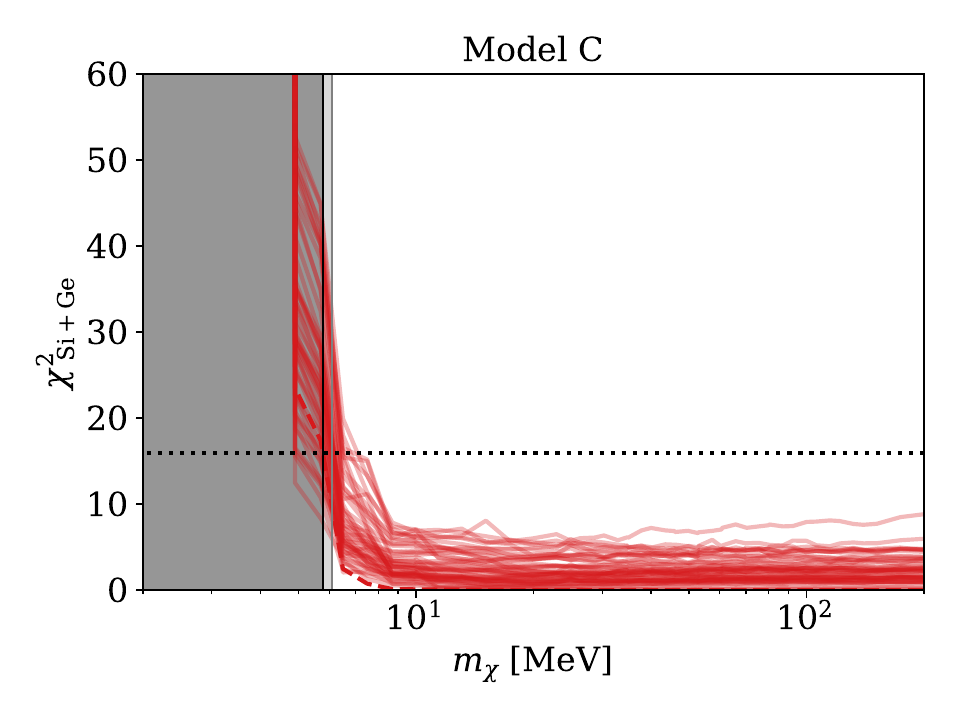}
		\includegraphics[width=0.495\textwidth]{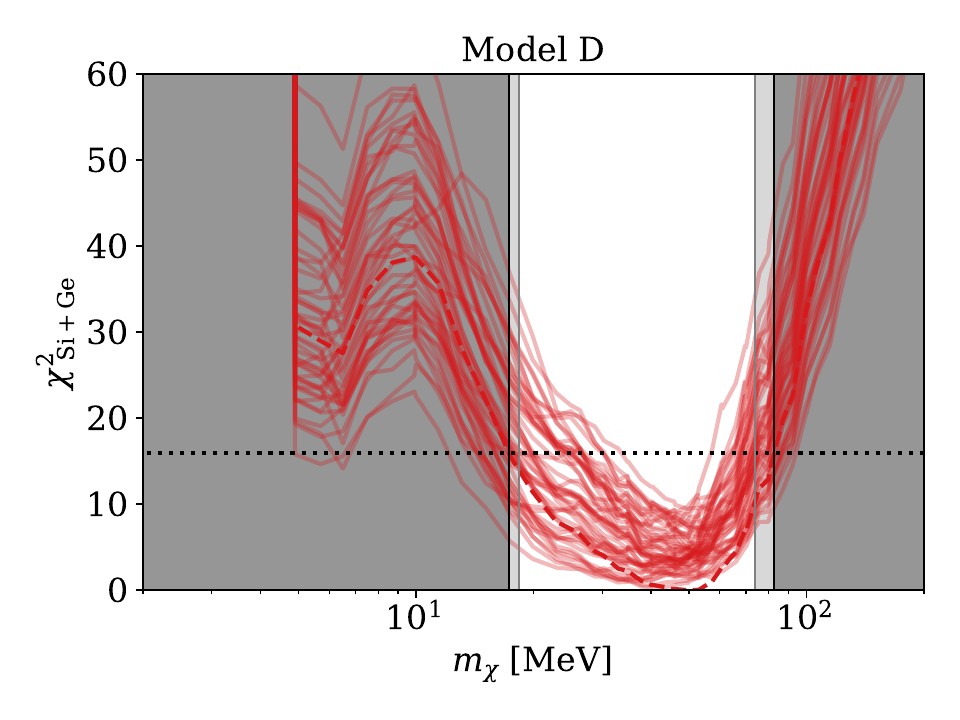}
		\caption{Joint Ge and Si $\chi^2$ for the best-fit velocity distribution as a function of the DM mass for the Asimov dataset (dashed red line) and 50 Poisson draws (solid red lines) at an exposure of 1~kg-yr. The dashed black line indicates the $\chi^2$ that corresponds to a 90\% CL exclusion limit for the relevant number of degrees of freedom (see table~\ref{tab:masslimits}). The dark gray shaded region indicates the associated excluded mass range based on the Asimov dataset. The light gray shaded region is the median excluded mass range derived from the Poisson draws.}
		\label{fig:masslimit_chisquared}
	\end{center}
\end{figure} 

So far we have discussed the best-fit $\chi^2$ only in terms of its qualitative behavior as a function of DM mass. In this section, we proceed to use it quantitatively as a test statistic in order to set limits on the DM mass based on our simulated data. Figure~\ref{fig:masslimit_chisquared} shows the $\chi^2$ curves of the Asimov\footnote{In the Asimov dataset \cite{Cowan:2010js} the number of observed events is exactly the number of expected events, i.e.~the Poisson distribution in the limit of infinite exposure.} 
datasets for models A--D (dashed red lines) together with the $\chi^2$ value that corresponds to a 90\% CL exclusion for each dataset (dotted black line). The number of degrees of freedom of the $\chi^2$ distribution underlying each of these thresholds is equal to the sum of the number of silicon bins and the number of germanium bins taken into account in the analysis. Here, we only use bins with an expected rate $\geq 5$ events to ensure that our test statistic is $\chi^2$-distributed to good approximation. The resulting numbers of degrees of freedom and the corresponding 90\% CL $\chi^2$ thresholds for each model are summarized in table~\ref{tab:masslimits}.

\begin{table}[thb]
    \centering
    \begin{tabular}{c||c|c|c|c|c}
        model &  d.o.f. (Si+Ge)\ & $\chi^2_{90\%}$ &$m_\chi$ [MeV] & $m_\chi$ lower limit [MeV] & $m_\chi$ upper limit [MeV]\\\hline
        A & 9+10 & 27.2 & 50 & 15 & -- \\
        B & 6+7 & 19.8 & 10 & 5 & 16 \\
        C & 5+5 & 16.0 & 50 & 6 & -- \\
        D & 5+5 & 16.0 & 50 & 17 & 83
    \end{tabular}
    \caption{Number of degrees of freedom relevant for setting a limit on $m_\chi$, associated 90\% CL $\chi^2$ threshold, and upper and lower bounds from the Asimov dataset at an exposure of 1~kg-yr for each model.}
    \label{tab:masslimits}
\end{table}

\begin{figure}[thbp]
	\begin{center}
		\includegraphics[width=0.7\textwidth]{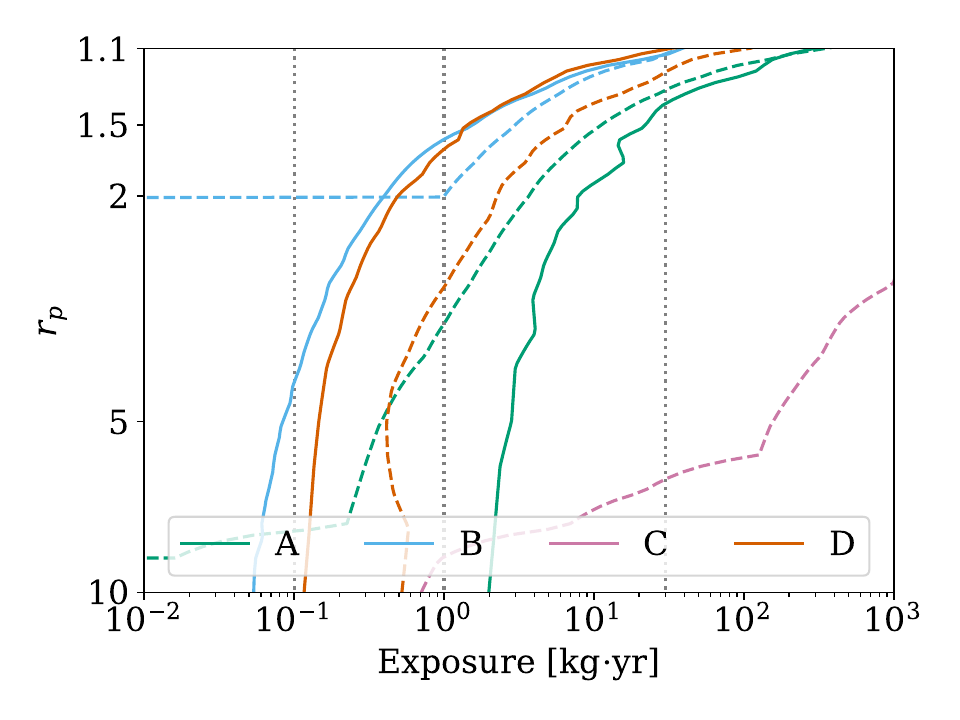}
		\caption{Relative precision $r_p$ for the upper bound (solid lines) and lower bound (dashed lines) on the mass at 90\% CL as a function of exposure for each model. The relative precision is defined as the ratio of the limit to the true mass (the true mass to the limit) for the upper (lower) bound (see main text for details). Future exposure targets of $0.1$~kg-yr, $1$~kg-yr and $30$~kg-yr are indicated by vertical dotted lines.}
		\label{fig:minimum_exposure}
	\end{center}
\end{figure} 

Using the best-fit $\chi^2$ curve as a function of $m_\chi$, we can translate the $\chi^2$ thresholds into 90\% CL exclusion limits on the DM mass. For the Asimov datasets of models A--D, this yields the exclusion shaded in dark gray in figure~\ref{fig:masslimit_chisquared}. The lower and upper bound of the allowed mass range for each model are given in the last column of table~\ref{tab:masslimits}.

In the Asimov dataset the number of observed events is exactly equal to the expected rate. To check how the mass limits obtained from this idealized construction compare to a more realistic scenario with Poisson fluctuations in the data, we generate 50 Poisson draws from the expected rates for each model. To this end, we draw event numbers for each model and experimental bin from the Poisson distribution with mean equal to the expected rate in the respective bin. We then derive mass limits based on each associated $\chi^2$ curve, which are all shown in figure~\ref{fig:masslimit_chisquared}. The median excluded mass range, corresponding to the median of the lower limits of the 50 draws and the median of their upper limits, is shaded in gray. For all of our models we find that the median exclusion from the Poisson draws is almost identical to the exclusion constructed from the Asimov dataset. This confirms that the latter is robust under statistical fluctuations and hence provides a good approximation to a realistic projected limit.

For models B and D we obtain an upper as well as a lower bound at 90\% CL from the Asimov dataset for the exposure of 1~kg-yr used in figure~\ref{fig:masslimit_chisquared}. For model A an approximately twice as large exposure is necessary to find an upper bound. This can be read off figure~\ref{fig:minimum_exposure}, which for each model shows the minimum exposure that yields a certain relative precision for the upper and lower limits on the DM mass. For the upper and lower bound we define the relative precision as $r_p \equiv m_\chi^\mathrm{max}/m_\chi$ and $r_p \equiv m_\chi/m_\chi^\mathrm{min}$, respectively, where $m_\chi^\mathrm{max}$ denotes the upper limit on the DM mass, $m_\chi^\mathrm{min}$ the lower limit and $m_\chi$ the true value. Larger exposures yield smaller values of $r_p$, i.e.\ more precise limits.

For the light-mediator model C no upper bound can be obtained for any exposure since arbitrarily large DM masses still yield a perfect fit to the Asimov dataset for some velocity distribution (see figure~\ref{fig:masslimit_chisquared}). For the other models there exists a minimum exposure above which an upper limit at a given confidence level can be found. In the case of a 90\% CL bound, this minimum exposure lies below the value of 1~kg-yr chosen for figure~\ref{fig:masslimit_chisquared} for models B and D, and above it for model A.

\section{Application to data}
\label{sec:senseivsedelweiss}

\begin{figure}[tbhp]
	\begin{center}
		\includegraphics[width=0.495\textwidth]{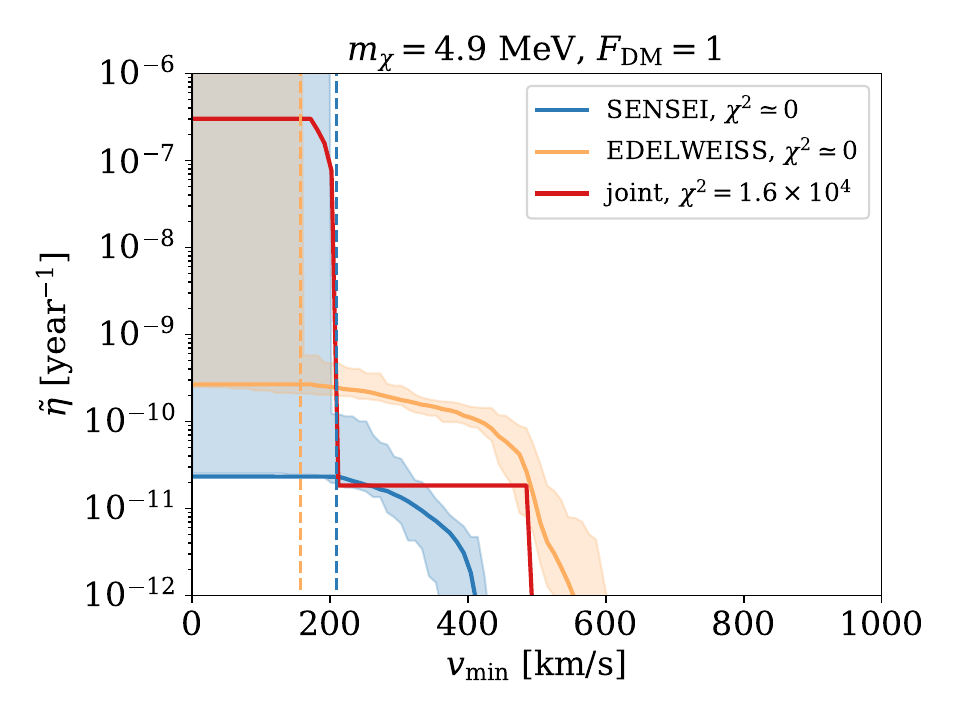}
		\includegraphics[width=0.495\textwidth]{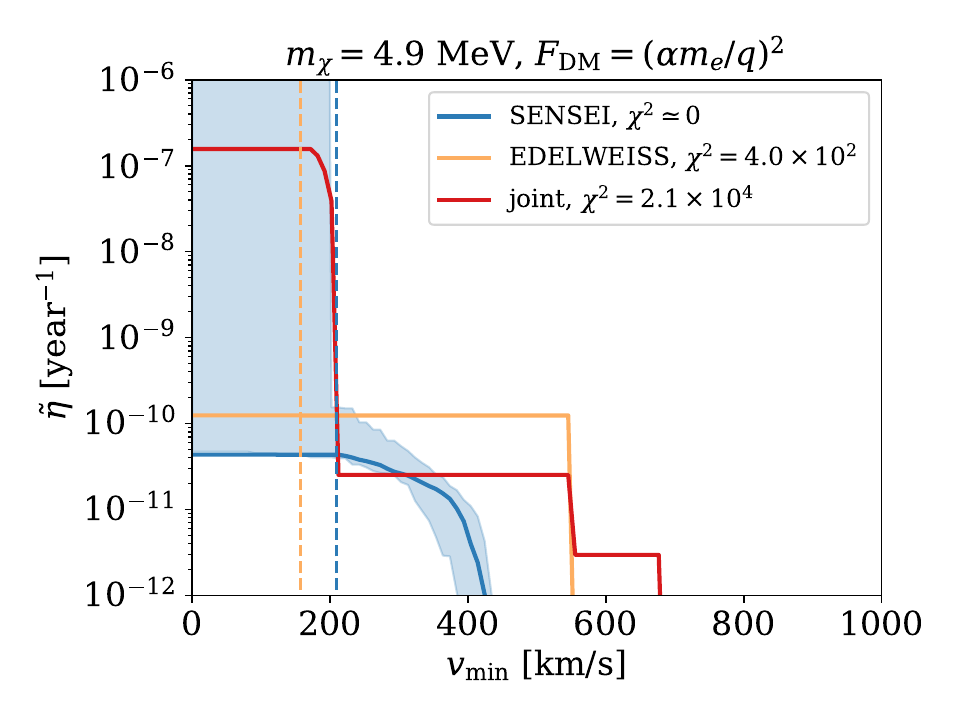}
		\includegraphics[width=0.495\textwidth]{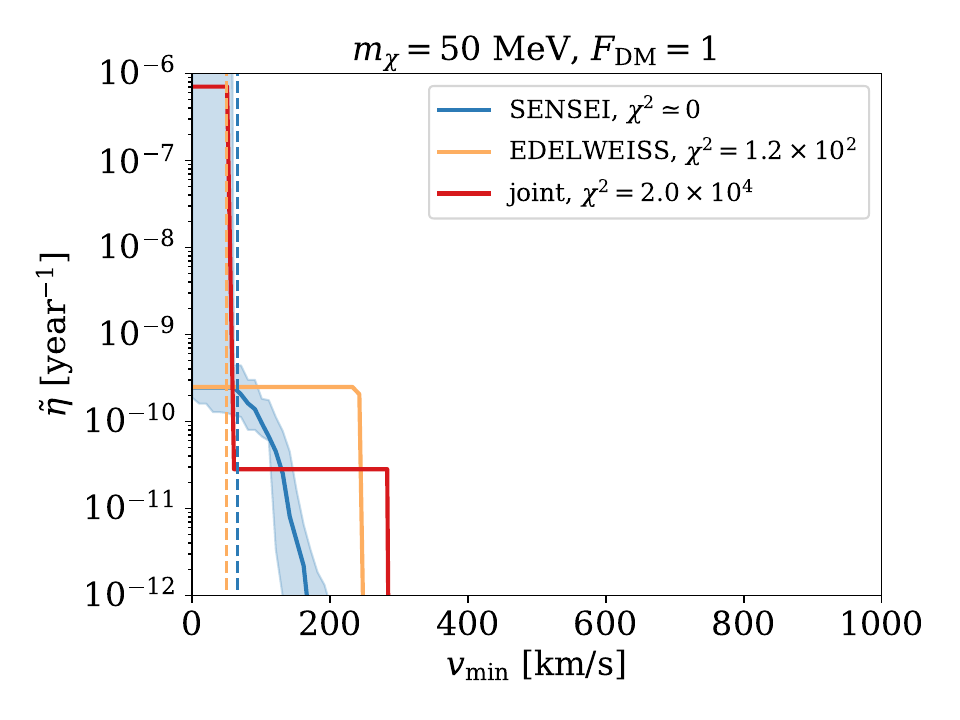}
		\includegraphics[width=0.495\textwidth]{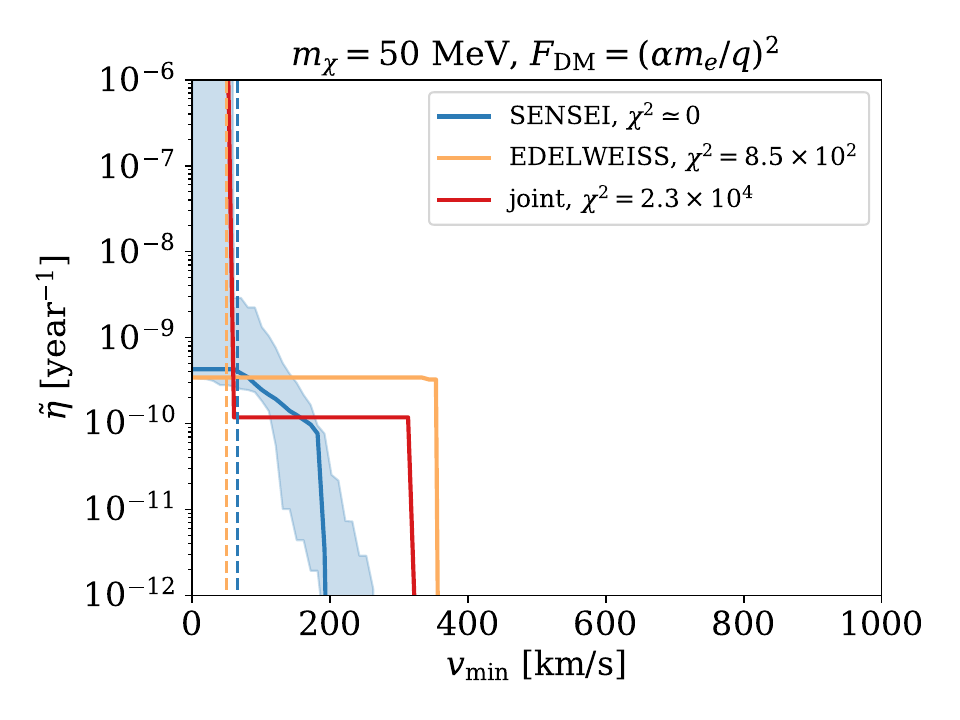}
		\caption{Best-fit velocity distributions inferred from fits of data from SENSEI, EDELWEISS, and both experiments simultaneously, shown for $m_\chi=4.9$~MeV (top) $m_\chi=50.0$~MeV (bottom) and for a heavy (left) or light (right) mediator. The bands denote the REST regions, as defined in section~\ref{sec:linalg}. We show these only for best fits with negligible $\chi^2$.}
		\label{fig:bestfit_velocity_data}
	\end{center}
\end{figure}

\begin{figure}[tbhp]
	\begin{center}
		\includegraphics[width=\textwidth, clip, trim = 0 37 0 0]{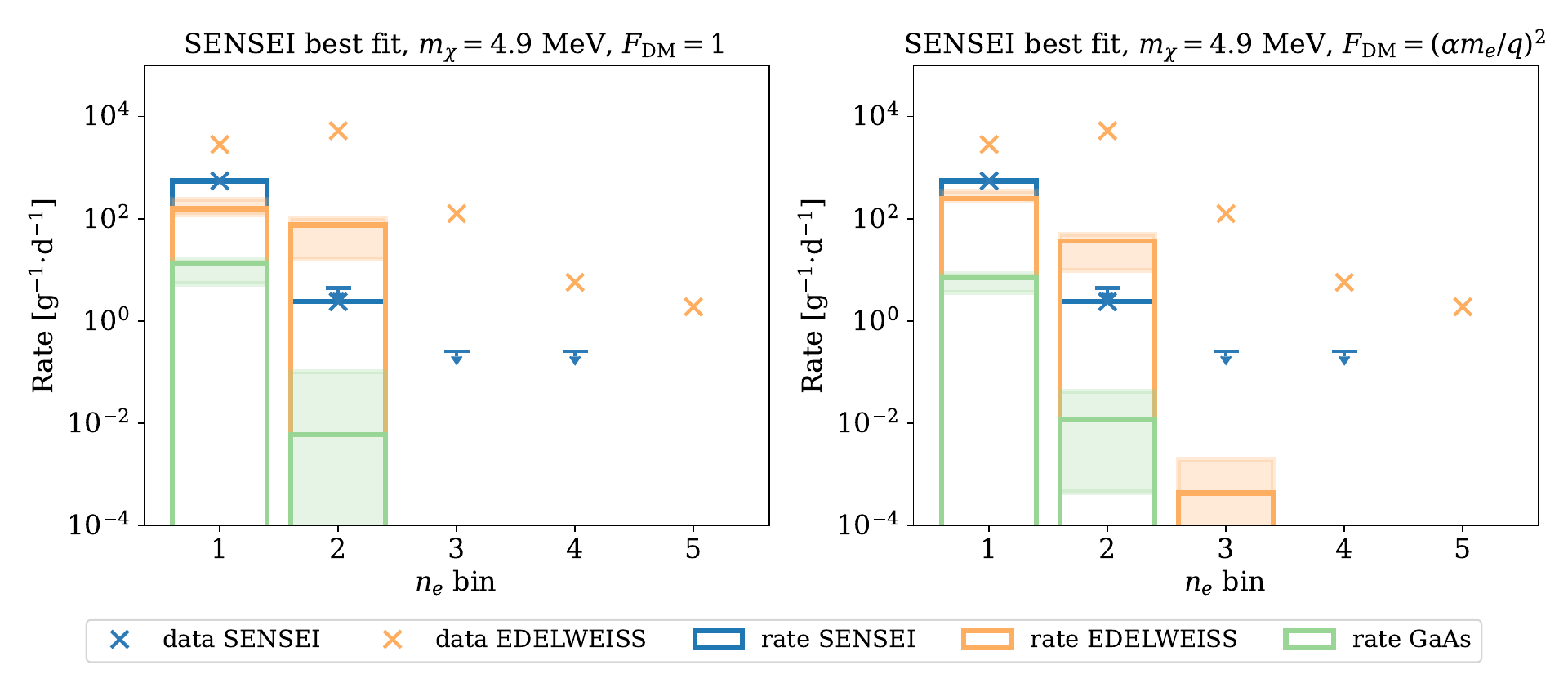}
        \includegraphics[width=\textwidth]{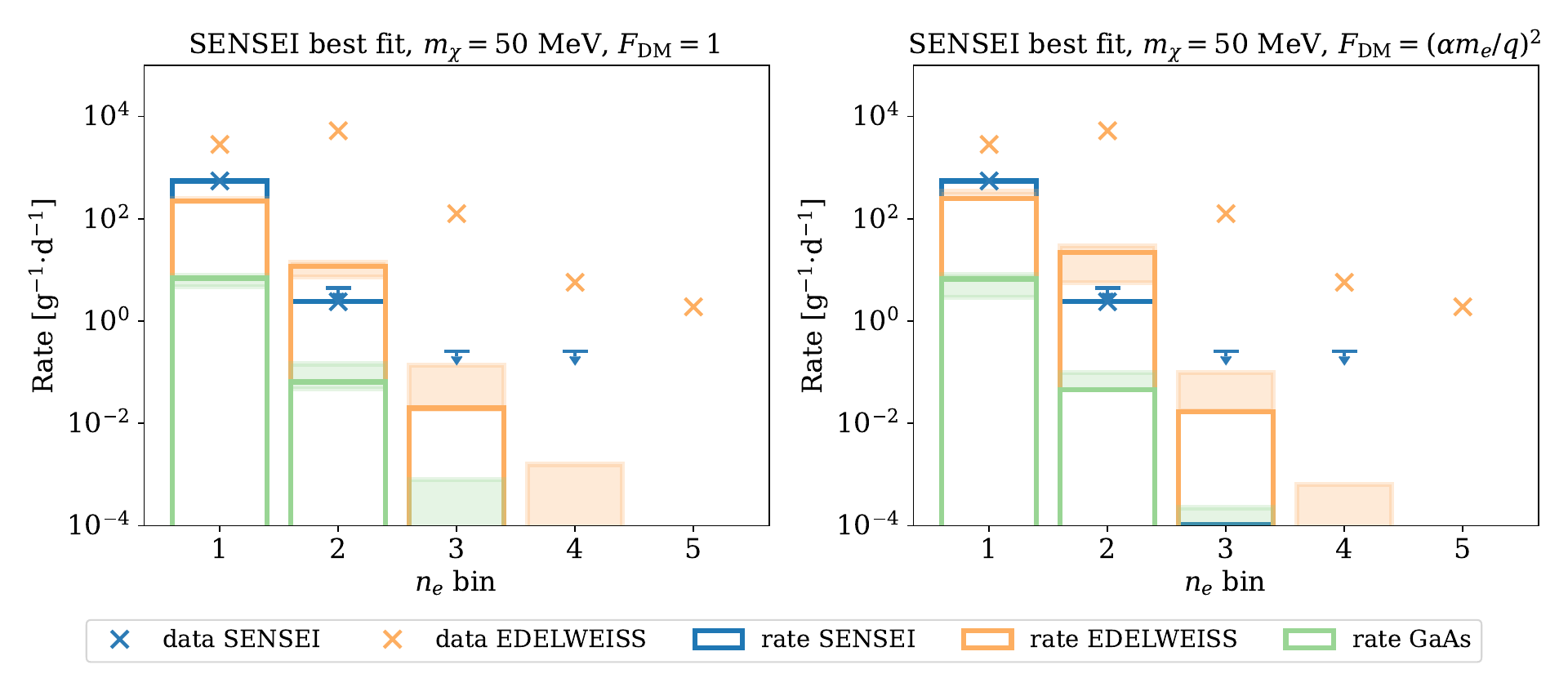}
		\caption{Predictions of the signal rates in EDELWEISS and a future GaAs detector based on the best-fit velocity distribution from SENSEI data, shown for $m_\chi = 4.9$~MeV (top) and $m_\chi = 50$~MeV (bottom) and for a heavy (left) or light (right) mediator. For the second, third and fourth SENSEI bin the 90\% CL upper limit on the rate is indicated with downward arrows. The shaded areas indicate the span of the predictions for EDELWEISS and GaAs from the REST band of the SENSEI fit.}
		\label{fig:senseibestfit_rates_data}
	\end{center}
\end{figure}

The mock-data based examples in the previous section provide a closure test which demonstrates that our procedure can reliably find the true dark matter model. We now apply our procedure to actual experimental data.  There are multiple experiments which have searched for DM-electron recoils: DAMIC \cite{DAMIC-M:2023gxo}, DarkSide-50 \cite{DarkSide:2022knj}, EDELWEISS \cite{EDELWEISS:2020fxc}, PandaX \cite{PandaX-II:2021nsg}, SENSEI \cite{SENSEI:2020dpa}, SuperCDMS \cite{SuperCDMS:2018mne}, XENON10 \cite{Essig:2017kqs}, and XENON1T \cite{XENON:2019gfn, XENON:2021qze}.  So far, none has found definitive evidence for DM-electron recoils.  Rather than carry out a comprehensive comparison across all experiments we illustrate our technique by focusing on two recent semiconductor-based searches, one using silicon (SENSEI) and one using germanium (EDELWEISS) as the target. However, our procedure can be easily expanded to include more searches.

SENSEI uses silicon Skipper CCDs to search for low-energy electron recoils~\cite{Tiffenberg:2017aac}. Silicon has a band gap of $1.2\eV$ and requires $3.8\eV$ for the creation of an additional electron-hole pair. In 2020, the collaboration took varying amounts of data up to the $4$-electron bin~\cite{SENSEI:2020dpa}. After adjusting for exposure and misclassification of $1\NE$ events the observed number of events in the $(1\NE, 2\NE, 3\NE, 4\NE)$ bins was $(1311.7, 5, 0, 0)$. After subtracting spurious-charge background events from the first bin, this translates\footnote{To be conservative, we subtract the $2\sigma$ lower bound on the spurious-charge background, $554$ events.  Note that to convert the number of observed events to a rate the exposure, which varies across electron bins, must be taken into account.} to a rate of $(549, 2.39, 0, 0)/(\mathrm{g}\cdot\mathrm{day})$.  Note that the properties of the materials used in experiments, \eg\ the band gap and the crystal form factor, are subject to systematic uncertainties. For simplicity, we do not vary these quantities within their uncertainties in the present demonstration of the halo-independent method, but a combined scan over material form factors and velocity distributions could be conducted efficiently using the vector space integration method of \cite{Lillard:2023nyj}

EDELWEISS uses germanium detectors utilizing Neganov-Trofimov-Luke amplification to go beyond electron-hole pair resolution~\cite{luke,Neganov:1985khw}.  Germanium has a somewhat smaller band gap and excitation energy than silicon, of $0.67\eV$ and $3\eV$, respectively.  In 2020, the EDELWEISS collaboration performed a search for DM-electron scattering using 33.4 g of germanium~\cite{EDELWEISS:2020fxc} and $58$ hours of data collection. In this search, the DM signal region extended to $30\eV_{ee}$, but here we only consider the data from the first $5$ electron bins.  Adjusting for the efficiency of signals to pass selection cuts and integrating the observed energy spectrum across each electron bin, we find a rate for electron recoils of $(1\NE,2\NE,3\NE,4\NE,5\NE)=
(2840, 5260, 126, 5.7, 1.9)$~events$/(\mathrm{g}\cdot\mathrm{day})$. Since there is presently not a model for possible backgrounds, we do not carry out any background subtraction.

There are hypothesized sources of background for these experiments, see \eg\ \cite{Du:2020ldo,SENSEI:2021hcn} for discussions of potential sources of single-electron events in Skipper-CCDs. Nevertheless, one can ask what DM velocity distribution is required if these rates are pure signal and whether they can be compatible with one another.

For the SENSEI dataset it is possible to find a physically viable velocity distribution that perfectly fits the observed data for any DM mass; for both a contact interaction $F_\text{DM}=1$ and a long-range interaction $F_\text{DM}\sim q^{-2}$.  EDELWEISS has a considerably larger event rate and has events out to higher energies than SENSEI.  Thus, their best-fit velocity distribution extends to higher speeds, and has a larger normalization. Moreover, only a heavy mediator ($F_\text{DM}=1$) and DM masses $m_\chi \lesssim 20$~MeV result in a good fit to the EDELWEISS data. For other DM hypotheses, a portion of the observed events has to originate from background, independent of the DM velocity distribution. Finally, when taken together, both experiments cannot be simultaneously fit for any choice of mass or coupling structure.

We illustrate these findings in figure~\ref{fig:bestfit_velocity_data}, where we show the best-fit velocity curves for SENSEI and EDELWEISS, as well as the joint fit for a selection of DM parameter choices. A DM mass of 4.9 MeV, shown in the top panels, provided the best joint fit, while a heavier mass of 50 MeV is shown for comparison in the lower panels. We see that even for the best-fit point of $m_\chi=4.9$ MeV, with the REST bands, the velocity distributions are incompatible, as demonstrated by the large $\chi^2$ value of $\chi^2=1.6~(2.1)\times 10^4$ for the heavy (light) mediator, respectively. Although we did not calculate beyond REST, in principle one can go beyond REST by following the procedure discussed in section~\ref{sec:linalg}. However, given the large size of the $\chi^2$ value in this example, going beyond REST does not provide any additional information.

Many of the best-fit velocity distributions in figure~\ref{fig:bestfit_velocity_data} are very different from the Standard Halo Model. In fact, only the $m_\chi = 4.9 \, \text{MeV}$, $F_\text{DM} = 1$ fit to the EDELWEISS data resembles the SHM (see models A--C in figure~\ref{fig:bestfit_velocity_asimov} for a comparison). With $m_\chi = 50\, \text{MeV}$, for example, the best fit to the SENSEI data is a co-rotating halo or disc with unexpectedly low dispersion. 
The power of the halo-independent approach is its ability to check the consistency of possible signals from multiple experiments, without imposing any prior for the distribution other than it be physical.  Thus, if the halo-independent approach finds two datasets to be inconsistent there is no physically allowed distribution that explains them both.

When comparing two experiments, there is another question one can ask; given a best-fit velocity profile for one set of experimental data, what should we expect in a different experiment? Specifically, we use the smaller velocity distribution required by SENSEI to predict a rate at EDELWEISS, under the (unlikely) assumption that SENSEI has no background events in their data. The expected rates at EDELWEISS for a DM particle of $4.9\MeV$ and $50\MeV$ are shown in figure~\ref{fig:senseibestfit_rates_data}, for two choices of the DM form-factor. The intervals between the lowest and the highest rate predicted by velocity distributions within the REST band of the fit to SENSEI data are shown as shaded areas in each bin. The large shortfall between the predicted EDELWEISS rate and the observed rate points to a substantial background. 

As a complementary exercise, we also show the predicted rate in a third, hypothetical experiment using GaAs. We see that the predicted rate for GaAs is lower than that of SENSEI. Thus, if this hypothetical GaAs experiment observes a higher rate than this prediction, then the experiment has a non-trivial background. On the other hand, if the GaAs experiment observes a lower rate, then this implies that the observed events in the SENSEI data are also contaminated by some background.

\section{Discussion and Conclusions}
\label{sec:conclusions}

The astrophysical DM-velocity halo distribution is a key ingredient to predicting the rate at dark matter direct detection experiments.  Equivalently, converting direct detection experimental data into insights about the particle physics properties of dark matter, such as its mass and couplings, depends upon assumptions about the astrophysics.  However, the halo model is subject to large uncertainties. These uncertainties propagate to the predictions for DM direct detection rates, which motivates a formalism in which one can present the results of DM direct detection experiments in a halo-independent manner.

In this work, we presented a procedure which extends the halo-independent formalism, previously applied to DM-nuclear recoils, to DM-electron scattering.  We use this approach to find the best-fit halo velocity distribution for a given set of data and DM parameters.  Profiling over these velocity distributions it is possible to determine the DM parameters that yield the best fit to the data, as well as exclude masses which do not fit the data for any physically-allowed velocity distribution.
Our procedure includes the Region of Equivalent Statistical Test (REST), which encapsulates the velocity models which report the same best-fit $\chi^2$ values and thus cannot be distinguished on the basis of the data (see section~\ref{sec:linalg}).  

We first performed a closure test in which we created mock data and then demonstrated that our procedure can accurately determine the true DM parameters and halo velocity distribution used to simulate the data.  Secondly, we applied our procedure to two current experiments, EDELWEISS and SENSEI,  to determine if there exists a viable DM model and halo velocity distribution which can simultaneously explain all the observed events at both experiments. 

From the closure test, we observed that 
the procedure works well for DM models with a heavy mediator, but is not as constraining for a light mediator. Rates from DM models with a light mediator are enhanced at low-momentum transfer, and therefore low energies.  As a consequence, they will only populate the lowest few energy bins.  In contrast, a heavy mediator will populate more energy bins, which provides more information for the analysis.  Generally, the efficacy of our halo-independent approach increases with the number of non-zero data bins.  For similar reasons our procedure provides tighter constraints as the dark matter mass increases.
  
The result of our halo-independent analysis of the SENSEI and EDELWEISS data indicates that the data cannot be explained by DM-electron scattering for \emph{any} DM velocity distribution if both experiments are free of background.  Furthermore, if the lower rate observed at SENSEI is due entirely to dark matter, EDELWEISS has a substantial background.  Finally, we predict that a GaAs experiment searching for DM-electron scattering~\cite{Derenzo:2016fse,TESSERACTTALK} should see a DM rate between $\sim 3$~events/(g-day) and $\sim 20$~events/(g-day), if SENSEI's rate is entirely due to dark matter.

Although we only analyzed data from semiconductor experiments, the technique outlined above is applicable to noble gas detectors, once the analogous ion form factors are calculated.
The rates in this work were generated using the {\tt QEDark} code. However, our procedure can be readily adapted to other codes, which can also compute $f_{\rm res}$ (\ref{eq:fcrystal}). The remainder of the procedure is unaffected by the choice of DM-electron scattering code. The procedure we outline parameterizes the velocity distribution as a piecewise-constant monotonic function with $\mathcal{O}(100)$ coefficients that must be determined.  Despite the large number of parameters, this minimization procedure is fast and takes only ${\cal O}(10{~\rm min})$ per model point. 

The halo-independent DM-electron scattering formalism presented here can also be extended to other types of DM models, such as inelastic DM, or to anisotropic detector materials. The formalism for inelastic DM-nuclear scattering was studied in~\cite{Bozorgnia:2013hsa,Scopel:2014kba}, where the authors extended the elastic nuclear recoil halo-independent formalism to account for the fact that the nuclear recoil energy does not uniquely map to $\vmin$, but instead has 2 solutions.
For anisotropic detector materials, the wavelet-harmonic method of \cite{Lillard:2023cyy} provides a suitable basis for the 3d velocity distributions and detector form factors, analogous to the linear basis of velocity functions in~(\ref{eq:etatilde_ansatz}).

If a robust DM signal is observed, the proposed formalism should be used to extract rigorous conclusions about the properties of DM. Additionally, it will be the only way to measure the local DM velocity distribution. The insights gained through this formalism could then inform the next experimental steps, including which other materials should be used to build follow-up detectors. 

\section*{Acknowledgments}
We thank Felix Kahlhoefer and Volodymyr Takhistov for helpful discussions. Fermilab is administered by Fermi Research
Alliance, LLC under Contract No. DE-AC02-07CH11359 with the U.S. Department of
Energy, Office of Science, Office of High Energy Physics.
The work of B.L. was supported in part by the U.S. Department of Energy under grant number DE-SC0011640. T.-T.Y. is supported in part by NSF CAREER grant PHY-1944826. 
T.-T.Y. also thanks the Center for Cosmology and Particle Physics at NYU for hospitality and support, where a portion of this work was completed.

\bibliographystyle{JHEP}
\bibliography{halo}
\begin{appendix}
\section{Halo Models}\label{app:halomodels}
Here, we present the analytic expressions for our two representative halo models, used in generating our mock data. 
\subsection*{Standard Halo Model}
In the galactic frame the Standard Halo Model is a Maxwell-Boltzmann distribution with a cutoff  
\be
f_\chi(\vec v_\chi)
=\frac{1}{K_{SHM}}e^{-v_\chi^2/v_0^2}\, \Theta(v_{\rm esc} -  v_\chi).
\ee
The normalization factor $K_{SHM}$ is determined by requiring $\int d^3 v f_\chi(\vec v)=1$, giving
\begin{eqnarray}
K_{SHM}&=& v_0^3\pi\left[ \sqrt{\pi} \textrm{Erf}\left(\frac{v_{\rm esc}}{v_0}\right)-2  \frac{v_{\rm esc}}{v_0} e^{-\left(\frac{v_{\rm esc}}{v_0}\right)^2}\right] \, .
\end{eqnarray}
\noindent We define the function $\eta(v_{\rm min})$, as in (\ref{eq:etadefinition}), 
\begin{equation}
\eta(v_{min}) = \int d^3 v \,\frac{e^{-(\vec v + \vec{v}_E)^2/v_0^2}}{v\, K_{SHM}}\,  \Theta(v - v_{\rm min}) \, \Theta(v_{\rm esc}-|\vec{v}+\vec{v}_E|)~.
\label{eq:etavmin}
\end{equation} 
The result of the integral takes a different form depending on whether $\vmin$ is bigger or smaller than $v_{\rm esc} - v_E$.
For $v_{\rm min}<v_{\rm esc}-v_E$,
\begin{eqnarray}
\eta(v_{\rm min})&=&\frac{v_0^2\pi}{2 v_E K_{SHM}} \left[-4e^{-v_{\rm esc}^2/v_0^2}v_E \right.\nonumber \\
 &&  \left.+\sqrt{\pi}v_0\left(\erf\left(\frac{v_{\rm min}+v_E}{v_0}\right)-\erf\left(\frac{v_{\rm min}-v_E}{v_0}\right)\right)\right]~,
\end{eqnarray}
while for $v_{\rm esc}-v_E < \vmin < v_{\rm esc}+v_E$
\begin{eqnarray}
\label{eq:etaSHM}
\eta(v_{\rm min})&=&\frac{v_0^2\pi}{2 v_E K_{SHM}}\left[
-2e^{-v_{\rm esc}^2/v_0^2}(v_{\rm esc}-v_{\rm min}+v_E) 
 \right. \nonumber \\
  && \left. + \sqrt{\pi}v_0\left(\erf\left(\frac{v_{\rm esc}}{v_0}\right)-\erf\left(\frac{v_{\rm min}-v_E}{v_0}\right)\right)\right]~.
 \end{eqnarray}

\subsection*{Stream}

We approximate the stream's velocity distribution as a narrow Gaussian, which in the Earth's frame takes the form
\be
  f(\vec{v})=\frac{1}{\sqrt{8\pi^3 }\sigma^3}e^{-\frac{(\vec{v} -\vec{v}_{\mathrm{str}})^2}{2\sigma^2}}\, 
  .
\ee
We are focused on the case where the stream's velocity is well below the galactic escape speed, $\vesc-|\vec{v}_{\mathrm{str}}-\vec{v}_E| \gg \sigma$, and it is a reasonable approximation to integrate over all speeds in the Earth's frame, \ie\ we do not include a Heaviside cutoff.
As before, the velocity distribution is normalized such that $\int d^3v f(v)=1$.  The function $\eta$ can be found by using (\ref{eq:etaSHM}) and making the substitutions $v_E\rightarrow v_{\mathrm{str}}$, $v_0\rightarrow \sqrt2 \sigma$ and taking the limit $\vesc \rightarrow \infty$ resulting in
\begin{equation}
\eta(\vmin) = \frac{1}{2v_{\mathrm{str}}}
\left(\erf\left(\frac{v_{\rm min}+v_{\mathrm{str}}}{\sqrt2\sigma}\right)-\erf\left(\frac{v_{\rm min}-v_{\mathrm{str}}}{\sqrt2\sigma}\right)\right)~.
\end{equation}

\section{The Pseudoinverse} \label{app:linearalgebra}

Consider an $(m\times n)$ matrix $A$ with rank $k$ whose, singular value decomposition is
\be
A = U D V^T~,
\ee
with $U$ an orthogonal $(m\times m)$ matrix, $V$ an orthogonal $(n\times n)$ matrix and $D$ a diagonal $(m\times n)$ matrix, of the form 
\be
D = \begin{pmatrix}
    D_k & 0 \\
    0 & 0
\end{pmatrix}~.
\ee
The rank $k$ square matrix $D_k$ is diagonal, with $(D_k)_{ii} \neq 0$. The singular values of $A$ are the diagonal entries of $D$, while $D_k$ includes only the nonzero singular values of $A$. By definition, $\det D_k \neq 0$. 
The pseudoinverse of $D$ is defined as 
\be
D^+ = \begin{pmatrix}
    D_k^{-1} & 0 \\
    0 & 0
\end{pmatrix}~.
\ee
The pseudoinverse of $A$ is defined in terms of $D^+$, 
\be
A^+ = V D^+ U^T~.
\ee
This pseudoinverse is the unique matrix \cite{penrose_1955} with the properties 
\begin{align}
A A^+ A &= A~,       &    A^+ A A^+ &= A^+ ~, \\
(A A^+)^T &= A A^+ ~,   &    (A^+ A)^T &= A^+ A~.
\end{align}
Furthermore, $A A^+$ acts as a projection operator, so that any vector can be projected into orthogonal image spaces using $A A^+$ and $\mathbb{1} - A A^+$, for example~$(\mathbb{1} - A A^+) x \in \ker(A)$.  

Finally, the least squares solution to the equation $A x = b$ is given by $x_0 = A^+ b$.  This can be seen by using the properties of the pseudoinverse to write 
\be
A x - b = A A^+ \left[A\left( x-A^+ b\right)\right] - \left(\mathbb{1} - A A^+\right)b~,
\ee
where we have used the projection operators to split the expression into pieces sitting in the two orthogonal spaces.  Thus, the norm of $A x - b$ is
\be
| Ax-b|^2 = |A\left(x-A^+ b \right) |^2 + |A x_0 - b |^2 \ge | A x_0 -b |^2~.
\ee
This shows that the pseudoinverse provides us with the least squares solution, and $x_0$ is the solution of minimum norm.  There may be other solutions, with larger norm, related to $x_0$ by an arbitrary vector in the kernel of $A$:
\be
x \rightarrow x_0 + \left(\mathbb{1} - A A^+\right) y~.
\ee
For the case of an underdetermined system ($n>m$) an exact solution $Ax_0 = b$ can always be found, provided the SVD of $A$ has at least $m$ nonzero singular values.  For an overdetermined system ($n<m$) there is no guarantee of an exact solution but $x_0$ is the least squares solution.

\end{appendix}
\end{document}